\documentclass[prb,longbibliography,twocolumn,superscriptaddress,floatfix,nofootinbib,showpacs,amsmath,amssymb]{revtex4-2}

\usepackage{graphicx}
\usepackage{dcolumn}
\usepackage{bm}
\usepackage{color}
\usepackage{tabularx}
\usepackage{csquotes,xcolor}
\usepackage[normalem]{ulem}
\usepackage{amssymb}
\usepackage{multirow}
\usepackage{subfiles}

\newcolumntype{Y}{>{\centering\arraybackslash}X}

\newcommand{\etal}{{\it et al.}}

\newcommand{\one}{{\bf 1}}
\newcommand{\erbium}{Er$^\mathrm{III}$}

\begin{document}


\title{Towards Understanding Prolate 4$f$ Monomers: Numerical Predictions and Experimental Validation of Electronic Properties and Slow Relaxation in a Muffin-shaped Er$^\mathrm{III}$ Complex}

\author{J.~Arneth$^+$}\email{jan.arneth@kip.uni-heidelberg.de}
\affiliation{Kirchhoff Institute for Physics, Heidelberg University, INF 227, D-69120 Heidelberg, Germany}
\author{C. Pachl$^+$}
\affiliation{Institute of Nanotechnology, Karlsruhe Institute of Technology (KIT),  Kaiserstr. 12, 76131 Karlsruhe, Germany}
\affiliation{Institute of Inorganic Chemistry, Karlsruhe Institute of Technology,  Kaiserstr. 12, 76131 Karlsruhe, Germany}
\author{G. Greif}
\affiliation{Institute of Inorganic Chemistry, Karlsruhe Institute of Technology, Kaiserstr. 12, 76131 Karlsruhe, Germany}
\author{B.~Beier}
\affiliation{Kirchhoff Institute for Physics, Heidelberg University, INF 227, D-69120 Heidelberg, Germany}
\author{P. W. Roesky}\email{roesky@kit.edu}
\affiliation{Institute of Inorganic Chemistry, Karlsruhe Institute of Technology,  Kaiserstr. 12, 76131 Karlsruhe, Germany}
\affiliation{Institute of Nanotechnology, Karlsruhe Institute of Technology (KIT),  Kaiserstr. 12, 76131 Karlsruhe, Germany}
\author{K. Fink}\email{karin.fink@kit.edu}
\affiliation{Institute of Nanotechnology, Karlsruhe Institute of Technology (KIT),  Kaiserstr. 12, 76131 Karlsruhe, Germany}
\author{R.~Klingeler}\email{klingeler@kip.uni-heidelberg.de}
\affiliation{Kirchhoff Institute for Physics, Heidelberg University, INF 227, D-69120 Heidelberg, Germany}

\begin{abstract}

We report the synthesis, crystal structure and magnetic properties of the triply-capped, slightly distorted trigonal-prismatic complex [Er(PPTMP)$_2$(H$_2$O)][OTf]$_3$ (PPTMP = (4-(6-(1,10-phenanthrolin-2-yl)pyridin-2-yl)-1H-1,2,3-triazol-1-yl)methyl pivalate) (\one). Complex \one\ is shown to exhibit field-induced slow relaxation of the magnetisation at $B = 0.1$~T via two distinct relaxation paths. Using tunable high-frequency/high-field electron paramagnetic resonance spectroscopy, we experimentally determine the effective $g$-factors and zero field splittings of the two energetically lowest Kramers doublets (KD). Our data reveal that the triply-capped, slightly distorted trigonal-prismatic ligand field favours an $m \simeq \pm 9/2$ magnetic ground state, while the main contribution to the first excited KD at $\Delta_{1 \rightarrow 2} = 780(5)$~GHz is suggested to be $m \simeq \pm 5/2$. The ground state $g$-tensor has generally axial form but hosts significant transversal components, which we conclude to be the source of SMM-silent behaviour in zero field. Our findings are backed up by \textit{ab-initio} spin-orbit configuration interaction calculations showing excellent agreement with the experimental data.

\end{abstract}
\maketitle
\def\thefootnote{+}\footnotetext{These authors contributed equally to this work.}\def\thefootnote{\arabic{footnote}}

\section{Introduction}

The seminal paper by Sessoli \etal~on the magnetic bistability in the famous dodecanuclear Mn$_{12}$ac-molecule~\cite{Sessoli1993} marked the birth of the ever-increasing research field centred around so-called single molecule magnets (SMM)~\cite{christou2000,sessoli2009,shao2020}. In SMM the combination of large spins and strong magnetic anisotropy leads to high effective energy barriers ($U_\mathrm{eff}$), due to which the magnetic moment relaxes only slowly and a remanent magnetisation can be retained over long timescales below the blocking temperature $T_\mathrm{B}$~\cite{winpennyBook}. Therefore, such complexes have potential applications in quantum computing~\cite{shiddiq2016}, high-density information storage~\cite{Coronado2019} and novel molecular spintronic devices~\cite{sanvito2010,Sanvito_2011}.

On the route to high-performance SMM $3d$ transition metal and $4f$ lanthanide ions have established themselves as the most promising building-blocks to realise high $U_\mathrm{eff}$ and $T_\mathrm{B}$~\cite{shao2020,Vieru2023}. While the large spin and magnetic anisotropy of polynuclear $3d$ coordination clusters typically arise from strong ferromagnetic interactions between the distinct magnetic centres, $4f$ moments exhibit good magnetic relaxation properties due to intrinsically strong spin-orbit coupling and large unquenched orbital momentum~\cite{sessoli2009,sorace2011,liddle2015}. Hence, almost 10 years after the discovery of Mn$_{12}$ac, Ishikawa \etal\ reported the first single ion magnet (SIM) [TbPc$_2$]$^-$, in which slow magnetic relaxation occurs for a single terbium(III) in a double-decker coordination~\cite{Ishikawa2003}. From there on, a vast amount of studies have been performed on $4f$-based SIM and record barriers of $U_\mathrm{eff} > 1800$~K~\cite{zhu2021} and magnetic hysteresis above liquid nitrogen temperatures $T_\mathrm{B} = 100$~K~\cite{EmersonKing2025} were achieved in this class of materials. However, despite huge zero field energy splittings, magnetic relaxation in SIM is often limited by under-barrier processes due to deviations from a perfectly axial crystal field. The resulting transversal magnetic anisotropy leads to a mixing of the $m_J$-levels, which enables quantum tunnelling of the magnetisation (QTM) between energetically degenerate states~\cite{thomas1996}. Therefore, the choice of a suitable coordination environment for the central $4f$ ion is a key step in the effective design of SIM.

In the literature, most of the reported SIM are based on the oblate $4f$ ions (Tb$^\mathrm{III}$, Dy$^\mathrm{III}$, Ho$^\mathrm{III}$), while those with predominantly prolate electron distribution (Er$^\mathrm{III}$, Tm$^\mathrm{III}$, Yb$^\mathrm{III}$) rather scarcely exhibit slow relaxation of the magnetisation~\cite{brown2015,gorczyski2015,zhang2018,castellanos2024,gavrikov2018,gugan2019,jain2024,Schwarz2025}. This mainly results from the need for equatorial ligand fields to minimize the Coulomb interaction with the paramagnetic ion and, hence, to enhance the single ion anisotropy, while still retaining a sufficient axiality~\cite{rinehart.2011}. However, this oblate versus prolate ion theory was demonstrated to be too simplified for the complex electronic structure of Er$^\mathrm{III}$, which renders trivalent erbium-based molecular complexes as interesting systems to investigate magnetic anisotropy in more detail~\cite{lucaccini2014,brown2015}. Nonetheless, although a wide variety of Er$^\mathrm{III}$ complexes with different relaxation properties,~i.e.~SMM, field-induced SMM and completely SMM-silent, are known, quantitative analysis of the single ion anisotropy is only rarely performed.

Here, we report the synthesis and magnetic characterisation of the triply-capped, slightly distorted trigonal-prismatic complex [Er(PPTMP)$_2$(H$_2$O)][OTf]$_3$ (PPTMP = (4-(6-(1,10-phenanthrolin-2-yl)pyridin-2-yl)-1H-1,2,3-triazol-1-yl)methyl pivalate) (\one ). Our ac susceptibility data show that \one\ exhibits field-induced slow relaxation of the magnetisation via two distinct relaxation pathways. We employ tunable high-frequency and high-field electron paramagnetic resonance (HF-EPR) spectroscopy which has been proven as a valuable method to study the magnetic anisotropy in Er$^\mathrm{III}$-based molecular complexes by direct experimental determination of the effective $g$-factors and zero field splittings of the low-energy Kramers doublets (KD)~\cite{spillecke2021,arneth2025}. The spectroscopic data indicate that SMM-silent behaviour in zero field arises from pronounced transverse components of the effective ground state $g$-tensor, while slow relaxation at finite fields occurs via an excited Kramers doublet at $\Delta_{1 \rightarrow 2} = 780$~GHz. A direct comparison of our experimental findings with the numerical results of spin-orbit configuration interaction calculations using the CASOCI program~\cite{bodenstein.2022} allows us to validate and assess \textit{ab-initio} predictions as recently done by us for various other mononuclear lanthanide complexes~\cite{comba2015,comba2018,spillecke2021,arneth2025}.

\section{Experimental Results}

\subsection{Structural Analysis}

\begin{figure}[b]
    \centering
    \includegraphics[width=\columnwidth]{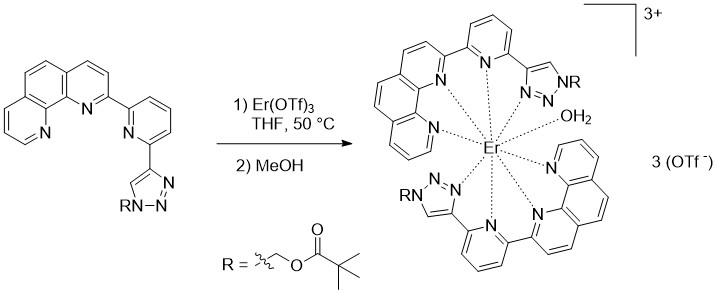}
    \caption{Synthesis scheme of complex \one .}
    \label{fig:synthesis_complex}
\end{figure}

\begin{figure}[h]
    \centering
    \includegraphics[width=\columnwidth]{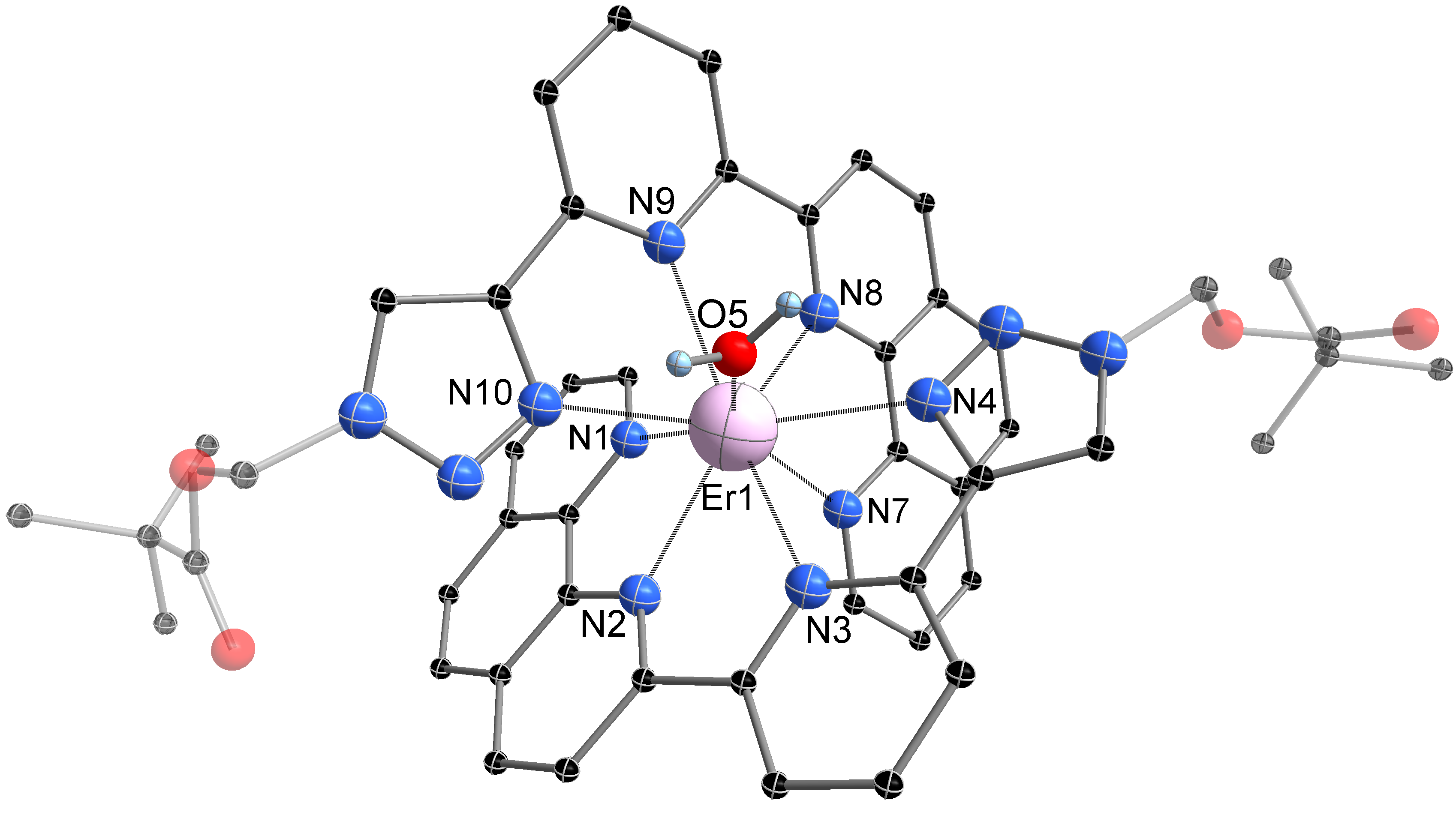}
    \caption{Molecular structure of [Er(PPTMP)$_2$(H$_2$O)][OTf]$_3$ in the solid state. C-H-bound hydrogen atoms are omitted for clarity.}
    \label{fig:structure}
\end{figure}

Reaction of PPTMP with erbium(III)-trifluoromethanesulfonate in THF led, after workup, to \one . Single crystals suitable for X-ray diffraction were grown from hot methanol (Scheme~\ref{fig:synthesis_complex}). Compound \one\ crystallises in the triclinic space group $P\bar{1}$ and contains two PPTMP-ligands coordinating the \erbium\ metal center. Fig.~\ref{fig:structure} shows the molecular structure of the erbium complex in the solid-state. The lanthanoid atom has a coordination number of nine, resulting from the coordination of eight nitrogen atoms from the two PPTMP ligands and one oxygen atom from a coordinated water molecule. This leads to a coordination geometry resembling a triply-capped, slightly distorted trigonal prism. In contrast, the three remaining triflate anions in the unit cell do not coordinate the lanthanide ion. 

In addition, the unit cell contains one non-coordinating methanol molecule, which interacts with the Er-coordinated water molecule via a hydrogen bond. Furthermore, two of the three triflate anions also exhibit hydrogen bonds: one hydrogen bond towards the coordinated water molecule, the other towards the methanol molecule. The water molecule originates from the non-dry solvents and the crystallisation was not carried out in the absence of moisture and oxygen. The bond distances between the erbium atom and the coordinating nitrogen atoms are in the range of 2.473(2) to 2.510(2)~\AA. As already observed in similar and phenanthroline-pyridine ligand systems~\cite{schnaubelt2017,petzold2017,petzold2018}, the nitrogen atoms N2 and N8 of the phenanthroline unit exhibit the shortest bond distances to the metal center with 2.473(2)~\AA\ each. The N–C bond lengths within the coordinated ligands also range from 1.327(3)~\AA\ to 1.373(3)~\AA, which is comparable with similar systems~\cite{schnaubelt2017,petzold2017,petzold2018}.

\subsection{Magnetometry}

\begin{figure}[h]
    \centering
    \includegraphics[width = \columnwidth]{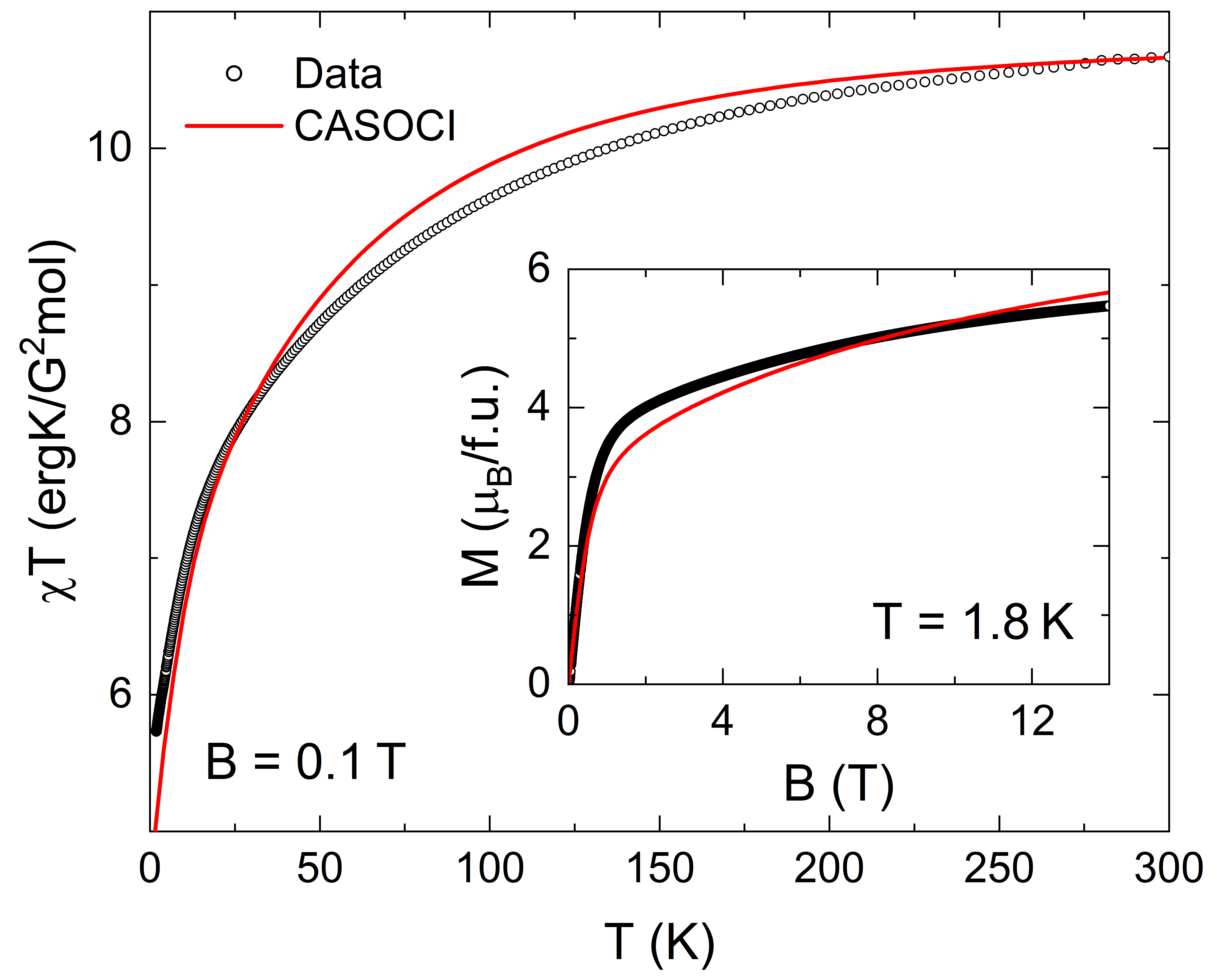}
    \caption{Temperature dependence of the $\chi T$-product at $B = 0.1$~T and isothermal magnetisation at $T = 1.8$~K (inset) of \one . Red lines depict the numerical results of \textit{ab-initio} CASOCI calculations as described in the text. }
    \label{fig:DCmag}
\end{figure}

The static magnetic susceptibility $\chi=M/B$ of \one\ in an external magnetic field of $B = 0.1$~T shown in Fig.~\ref{fig:DCmag} by its $\chi T(T)$-profile is typical of mononuclear Er$^\mathrm{III}$ complexes~\cite{rechkemmer2015,zhang2018,bazhenova2021}. At $300$~K, the measured value of $\chi T = 10.71\,\mathrm{ergK/G^2mol}$ is close to $\chi T_\mathrm{free} = 11.48\,\mathrm{ergK/G^2mol}$ expected for a free Er$^\mathrm{III}$ ion. Since a sizeable depletion due to crystal field effects is ruled out by the numerical calculations, as described later, the remaining discrepancy likely arises from solvent molecules cocrystallising in the packing structure. Upon cooling, $\chi T$ gradually decreases and then rapidly drops below $30$~K, which is attributed to the depopulation of the Stark levels. The isothermal magnetisation at $T = 1.8$~K (inset of Fig.~\ref{fig:DCmag}) is characterised by a steep increase at low fields followed by a plateau-like behaviour above $1$~T. Up to the highest accessible field of $14$~T the magnetisation gradually increases to $M(14~{\rm T}, 1.8~{\rm K}) = 5.5\,\mathrm{\mu_B/f.u.}$ (formula unit), i.e., does not reach the full saturation value ($M_\mathrm{sat} = 9\,\mathrm{\mu_B/f.u.}$), thereby confirming the presence of considerable magnetic anisotropy.

\begin{figure}[htb]
    \centering
    \includegraphics[width = \columnwidth]{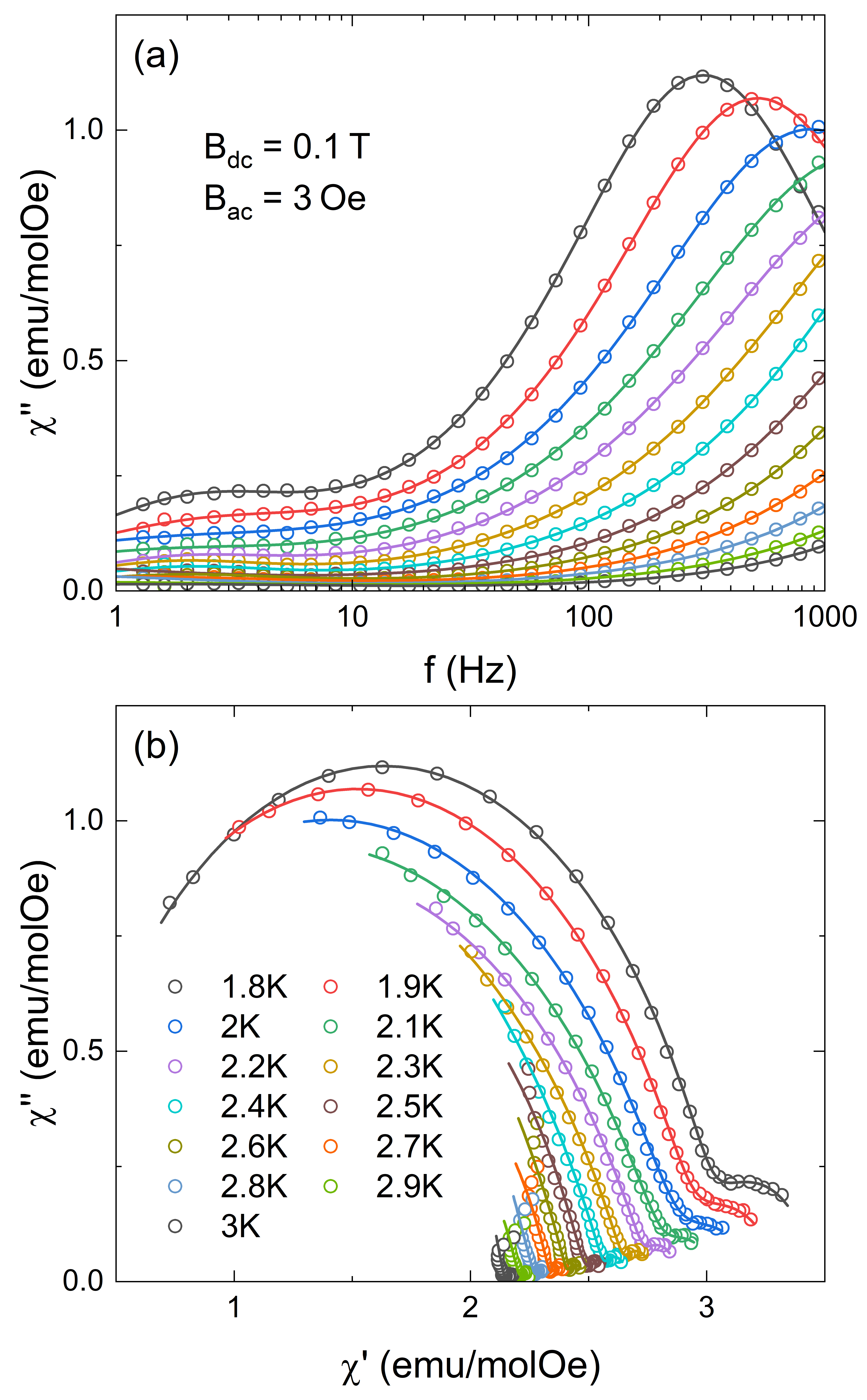}
    \caption{Out-of-phase ac susceptibility (a) and Cole-Cole plot (b) of \one\ at selected temperatures and external static magnetic field $B_\mathrm{dc} = 0.1$~T. Solid lines depict fits to the data using eq.~\ref{eq:doubleDebye} as described in the text.}
    \label{fig:ACChi}
\end{figure}

While complex \one\ does not exhibit slow magnetic relaxation (up to $1$~kHz) at $B = 0$~T, a clear out-of-phase ac susceptibility signal is found at low temperatures when a small static external field of $B_\mathrm{dc} = 0.1$~T is applied (Fig.~\ref{fig:ACChi}). From the frequency dependence of $\chi{''}$ (a), two maxima can be clearly discerned as confirmed by the appearance of two distinct lobes in the Cole-Cole plot (b). Hence, the magnetisation in \one\ relaxes via two separate pathways with differing relaxation rates. To corroborate our assignment, the ac susceptibility data were fitted by the sum of two generalised Debye-functions:

\begin{equation}
    \chi_\mathrm{ac}(f) = \sum_{n = 1}^2 \left( \chi_{\rm{S},n} + \frac{\chi_{\rm{T},n} - \chi_{\rm{S},n}}{1 + (if\tau_n)^{1-\alpha_n}} \right),
    \label{eq:doubleDebye}
\end{equation}

where $\chi_\mathrm{S}$ and $\chi_\mathrm{T}$ denote the adiabatic and isothermal susceptibilities, $\tau$ is the relaxation time, and $0 \leq \alpha \leq 1$ qualitatively represents the distribution of relaxation times within the powder sample. As visible in Fig.~\ref{fig:ACChi}, the data are well described by the used model. Although the limited frequency range does not allow for a quantitative analysis, both relaxation processes are qualitatively found to become faster upon increasing the temperature. The absence of SMM-behaviour in zero field straightforwardly implies a considerable transversal magnetic anisotropy acting on the \erbium\ moment. Such off-diagonal elements in the single ion anisotropy tensor lead to a mixing of the $m_J$-states and, thus, give rise to fast QTM, which can be effectively quenched by an external magnetic field~\cite{thomas1996}.

\subsection{High-field EPR studies}

\begin{figure}[t]
    \centering
    \includegraphics[width = \columnwidth]{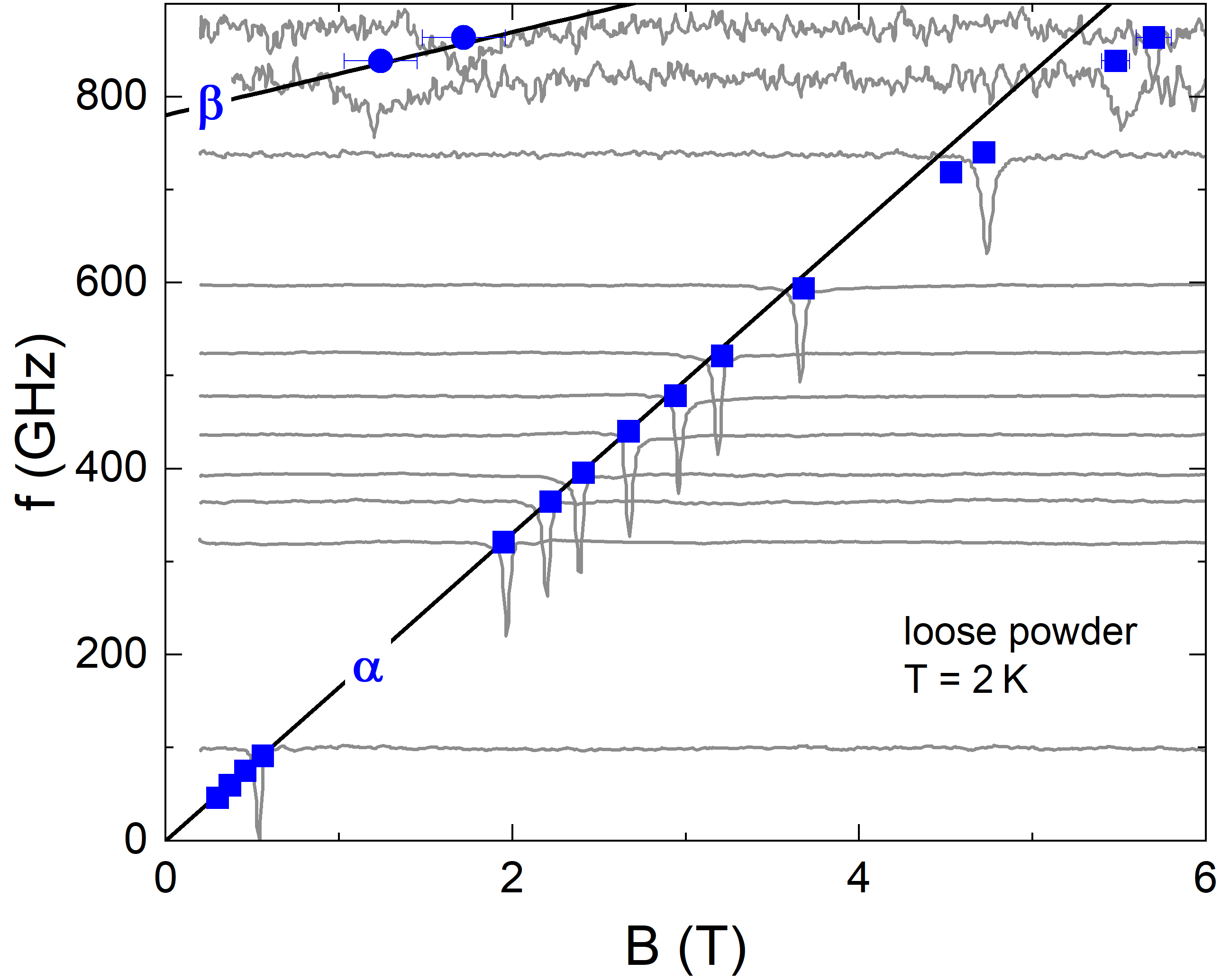}
    \caption{Resonance frequency versus magnetic field diagram for a loose powder sample of \one\ obtained at $T = 2$~K. Blue squares mark the resonance positions as read off the measured HF-EPR spectra displayed in the background. Solid black lines depict simulations of the ground state resonance branches using a $S = 1/2$ pseudospin model for each KD as described in the text.}
    \label{fig:fB2K}
\end{figure}

To experimentally probe the low-energy magnetic excitations in \one\ and, in particular, to quantitatively determine the magnetic anisotropy, we performed multifrequency HF-EPR studies both on loose and fixed powders of freshly ground samples. The oriented loose-powder HF-EPR spectra obtained at $T = 2$~K (Fig.~\ref{fig:fB2K}) display a pronounced Lorentzian-shaped absorption feature in the entire accessible frequency range up to $900$~GHz. For frequencies higher than ca.~$800$~GHz, an additional weak broad resonance feature appears at smaller magnetic fields. The extracted resonance field positions at different frequencies are shown in Fig.~\ref{fig:fB2K}, too. As visualised by the solid black lines, the pronounced main feature forms a gapless resonance branch, here labelled as $\alpha$, which exhibits linear field dependence at small magnetic fields but begins to deviate from its initial linear behaviour for $B \gtrsim 4$~T. In agreement with the findings of the ac susceptibility studies and as further discussed below, the high-field bending of resonance branch $\alpha$ is associated with an avoided level crossing and, hence, indicates the mixing of different $m_J$-levels. Linear \mbox{extrapolation} of the two data points marking the weak low-field feature yields a tentative resonance branch $\beta$ with a zero field splitting (ZFS) of $\Delta \simeq 780$~GHz. From the slopes of the observed resonance branches, we read off the effective $g$-factors $g_\mathrm{eff} \simeq 11.8$ and $3.2$ for $\alpha$ and $\beta$, respectively.

\begin{figure}[t]
    \centering
    \includegraphics[width = \columnwidth]{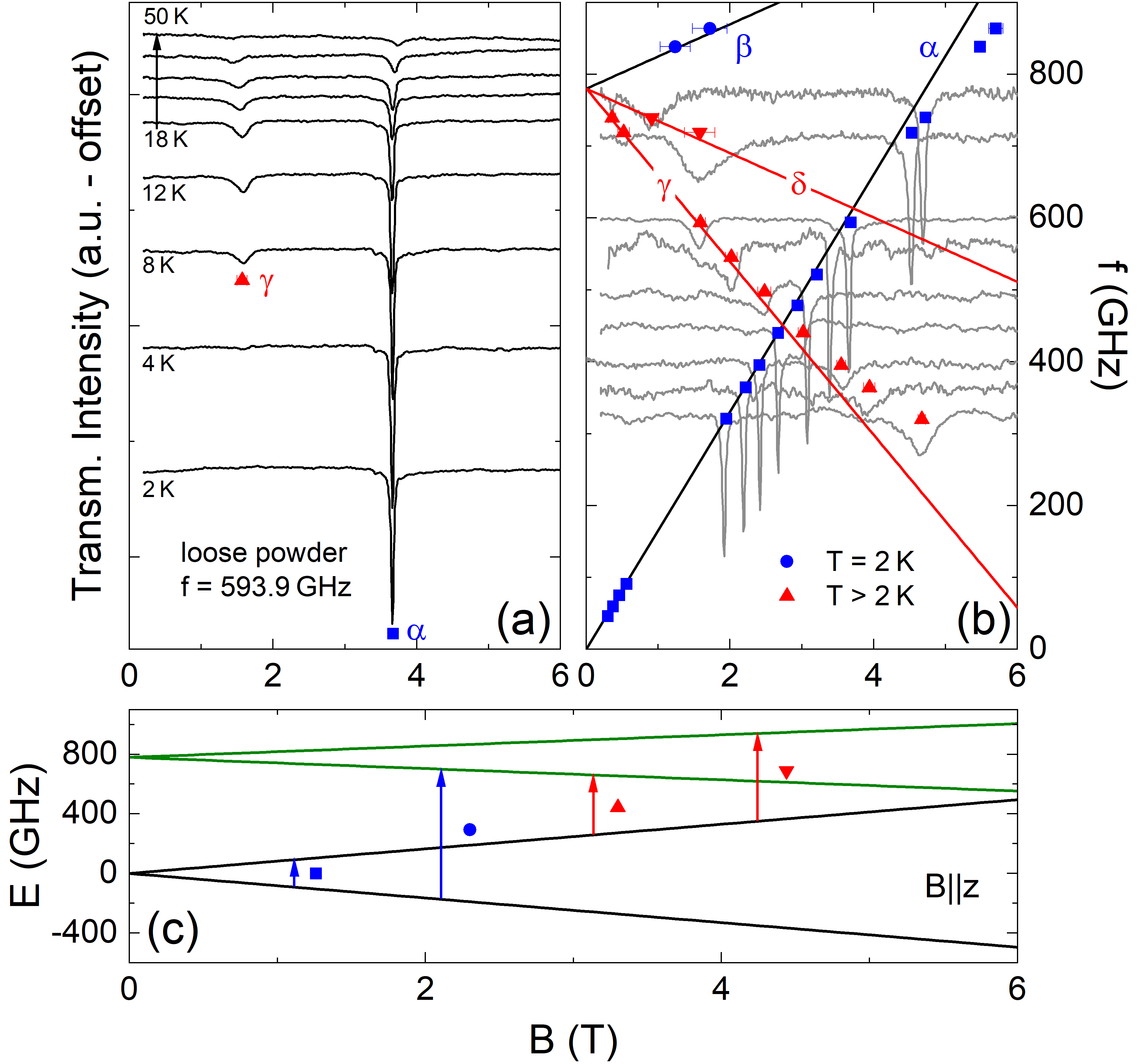}
    \caption{Loose powder HF-EPR spectra (shifted along the ordinate) of \one\ at selected temperatures and at fixed frequency $f = 593.9$~GHz (a), the corresponding frequency-field diagram (b) and simulated energy levels of the two energetically-lowest KDs (c). Solid black/red lines in (b) depict simulated ground/excited state transitions corresponding to the arrows in (c) which follow the same color scheme. Grey lines in the background of (b) display HF-EPR spectra measured at $T = 18$~K. }
    \label{fig:highT}
\end{figure}

The effect of temperature on the HF-EPR spectra is shown in Fig.~\ref{fig:highT}(a) at an exemplary frequency of $f = 593.9$~GHz. Upon heating, the spectral weight of the pronounced main feature corresponding to the resonance branch $\alpha$ gradually decreases until it becomes almost indiscernible around $50$~K. Further, starting from $T \simeq 8$~K, an additional absorption peak ($\gamma$) appears in the spectra, which first increases in intensity with rising temperature and then vanishes again for $T \geq 50$~K. To investigate the field dependence of $\gamma$, HF-EPR spectra at different frequencies were acquired at $T = 18$~K, i.e., where the resonance feature is most pronounced. As visible in the spectra and the corresponding frequency-field diagram in Fig.~\ref{fig:highT}(b), two additional resonance branches ($\gamma$ and $\delta$) are observed at elevated temperatures. Similar to the behaviour of $\alpha$, $\gamma$ follows a linear field dependence at small fields and exhibits significant bending at $B \gtrsim 4$~T. As will be discussed in detail below, this non-linear field dependence further confirms the occurrence of state mixing in \one . The zero field gaps of $\gamma$ and $\delta$, obtained by extrapolating the branches to $B = 0$~T, coincide with that of $\beta$ within the error bars of our experiment.

Due to the absence of a zero field excitation gap and its Curie-like temperature dependence, it is straightforward to attribute $\alpha$ to a transition within the energetically lowest Kramers doublet (KD1). In contrast, the observed finite ZFS of branches $\beta$-$\delta$ render them likely to be associated with transitions between two distinct KDs. While the occurrence of $\beta$ at $T = 2$~K implies the magnetic ground state (GS) to be its initial state, the temperature dependencies of $\gamma$ and $\delta$ show that the corresponding transitions arise from thermally excited states (ES) a few K above the GS.

The observed resonance branches can be rationalised by assuming an $S = 1/2$ pseudospin model for the two energetically lowest KDs. The best simulations (solid lines in Fig.~\ref{fig:highT}(b)) are obtained using a zero field splitting of $\Delta_\mathrm{1 \rightarrow 2} = 780(5)$~GHz and effective $g$-factors of $g_\mathrm{eff,1} = 11.8(2)$ and $g_\mathrm{eff,2} = 5.4(3)$ for KD1 and KD2, respectively. As can be seen from the corresponding simulated Zeeman diagram in Fig.~\ref{fig:highT}(c), the upwards branch of KD1 and the downwards branch of KD2 cross at $B \simeq 6$~T. While transverse components of the magnetic anisotropy tensor are effectively neglected in the $S = 1/2$ pseudospin model, their presence in the actual complex is clearly suggested by an avoided-level-crossing behaviour which qualitatively explains the observed bending of branches $\alpha$ and $\gamma$.

\begin{figure}[t]
    \centering
    \includegraphics[width = \columnwidth]{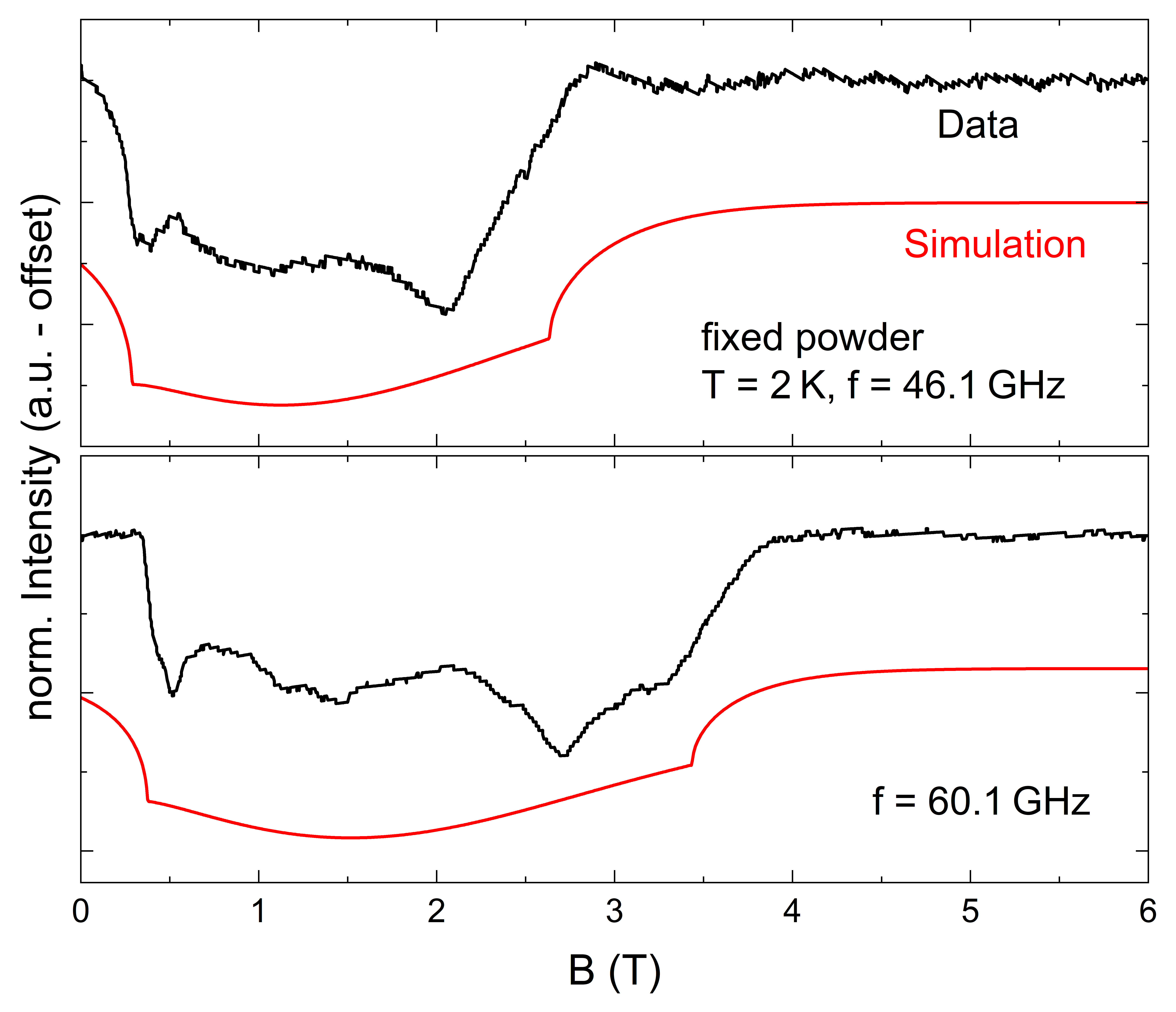}
    \caption{Measured (black) and simulated (red) fixed powder HF-EPR spectra of \one\ at $T = 2$~K for $f = 46.1$~GHz (top) and $f = 60.1$~GHz (bottom). }
    \label{fig:fixed}
\end{figure}

Due to the alignment of the sample with the external magnetic field, the loose powder studies provide quantitative information only on the easy anisotropy direction of the studied complex. To determine also the transverse components of the effective $g$-tensor, we performed HF-EPR measurements on a fixed powder of \one . Fig.~\ref{fig:fixed} displays the corresponding spectra obtained at $T = 2$~K for two exemplary frequencies in the V-band range. The general shape of the low-temperature fixed powder HF-EPR spectra is given by a broad absorption minimum superimposed by several sharper (but in comparison to the loose powder spectra very weak) resonance features. While the former is a typical characteristic of the measured powder average, the latter likely arise from clusters of crystallites that remained partially ordered after the loose powder measurements. The wide distribution of spectral weight straightforwardly indicates considerable transversal components of $g_\mathrm{eff}$ for KD1, i.e., a mixing of states due to off-diagonal crystal field parameters, which elucidates the absence of SMM behaviour in \one\ at $B = 0$~T. The components of $g_\mathrm{eff}$ can be quantified by fitting the overall shape of the spectra using a $S = 1/2$ pseudospin model with a strongly anisotropic $g$-tensor. The best agreement with the experimental data is achieved by $g_\mathrm{eff,x} = 1.2(3)$, $g_\mathrm{eff,y} = 2.6(4)$ and $g_\mathrm{eff,z} = 11.8(3)$ and the simulated spectra are shown as red lines in Fig.~\ref{fig:fixed}.

\subsection{Quantum Chemical Calculations}

\begin{figure}[t]
    \centering
    \includegraphics[width = \columnwidth]{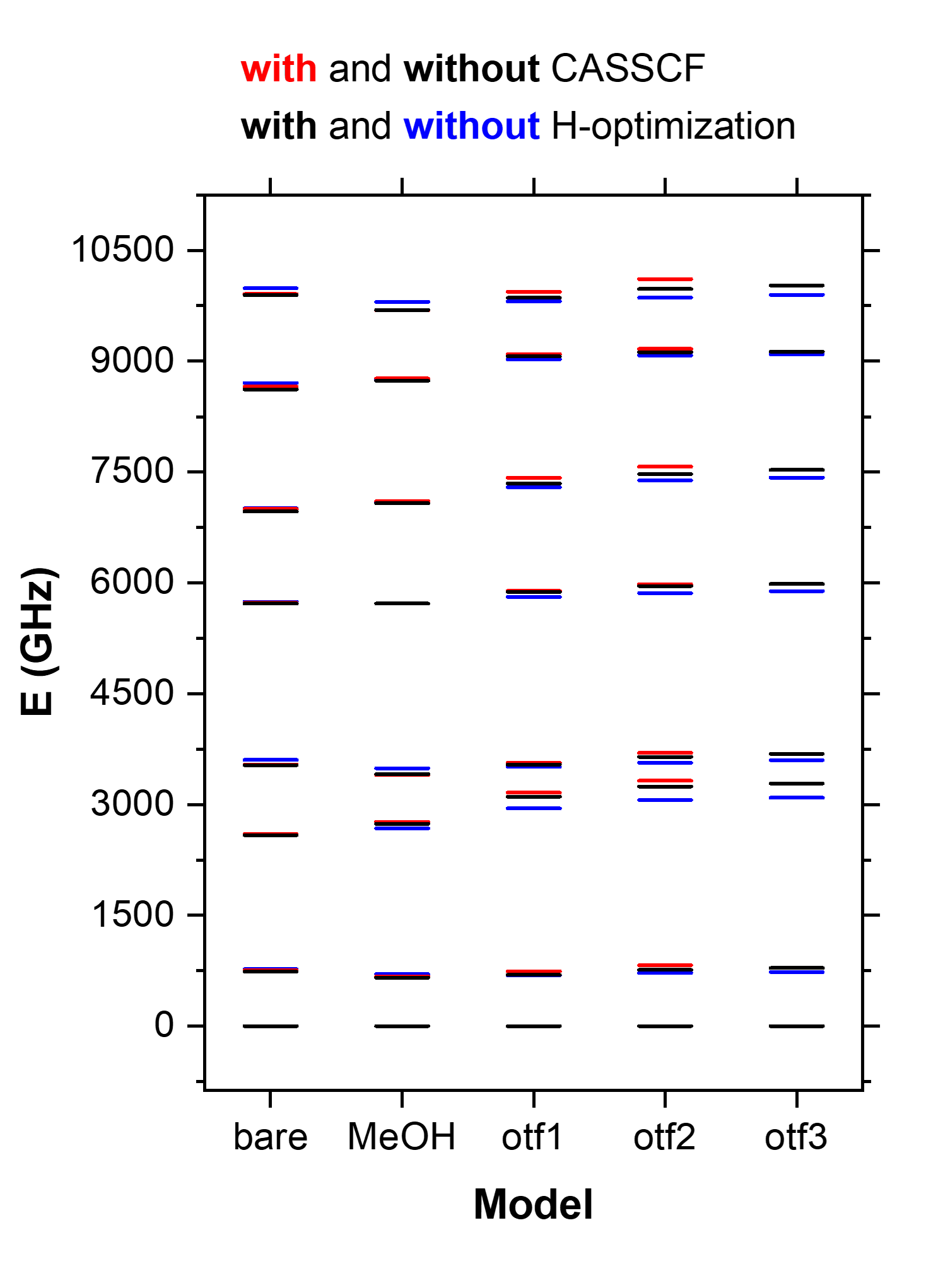}
    \caption{Calculated Energies of the 8 lowest Kramers Doublets of the $^4I_{15/2}$-ground state of complex \one. \textbf{Black}: Energies with H-optimization and without intermediate CASSCF step. \textbf{\color{blue}Blue}: Energies without H-optimisation and without intermediate CASSCF step. \textbf{\color{red}Red}: Energies with H-optimisation and intermediate CASSCF step.}
    \label{fig:splitting1}
\end{figure}

Using the workflow described in the methods section, the influence of the counterions on the magnetic properties is investigated. Therefore, five model clusters are simulated: The first is the bare molecule with a charge of +3. Secondly, the MeOH hydrogen bound to the coordinating water is included, which remains in the charge state +3. Next, the three counterions are added sequentially. The model complexes, therefore, exhibit a charge of +2, +1 and 0, respectively. Pictures of the different model complexes can be found in the SI, figure S10. Furthermore, calculations are carried out with and without a preliminary optimisation of hydrogen positions. Lastly, an intermediate CASSCF-step is tested for H-optimised models. This calculation has not been performed for the largest model, due to an internal basis set limit. The resulting splitting of the $^4I_{15/2}$-ground state is visualised in Fig.~\ref{fig:splitting1}. The exact values, diagonal elements of the $g$-tensors for each KD, extended Stevens operator equivalents, and $m_J$-compositions for all calculations are summarised in the Supplemental Material, tables II-XXXIII.

As can be seen, neither the optimisation of the hydrogen positions nor the inclusion of an intermediate CASSCF-step significantly influence the calculated energy splittings. The observed differences between the respective models are smaller than $120$~GHz for the first excited KD. Therefore, the optimised calculations without an intermediate CASSCF-step are used as the reference. The influence of including the counterions is only marginally larger, as the lowest value of $\Delta_{1 \rightarrow 2} \simeq 650$~GHz is calculated for the model that includes MeOH and $790$~GHz is found for the full model. The small differences in energy translate to weak discrepancies in the calculated $g_z$-values of the ground state KD, which are between 11.8 and 12.8 for all calculated models. A similar difference in absolute values can be reported for the transversal components of the $g$-tensor, with values ranging from 0.10 to 1.63 for $g_\mathrm{x}$ and from 0.89 to 2.71 for $g_\mathrm{y}$. Thus, the H-optimisation is shown to notably influence the transversal components of the $g$-tensor. Not optimising hydrogen positions and calculating the bare cluster would significantly overestimate the axiality of the underlying model by minimising the transversal components. Considering the optimised H positions and including also the counterions yields transversal $g$-tensor components of $g_\mathrm{x} = 1.10$ and $g_\mathrm{y} = 1.95$, which are in very good agreement with the experimental results.

\begin{figure}[bht]
    \centering
    \includegraphics[width = \columnwidth]{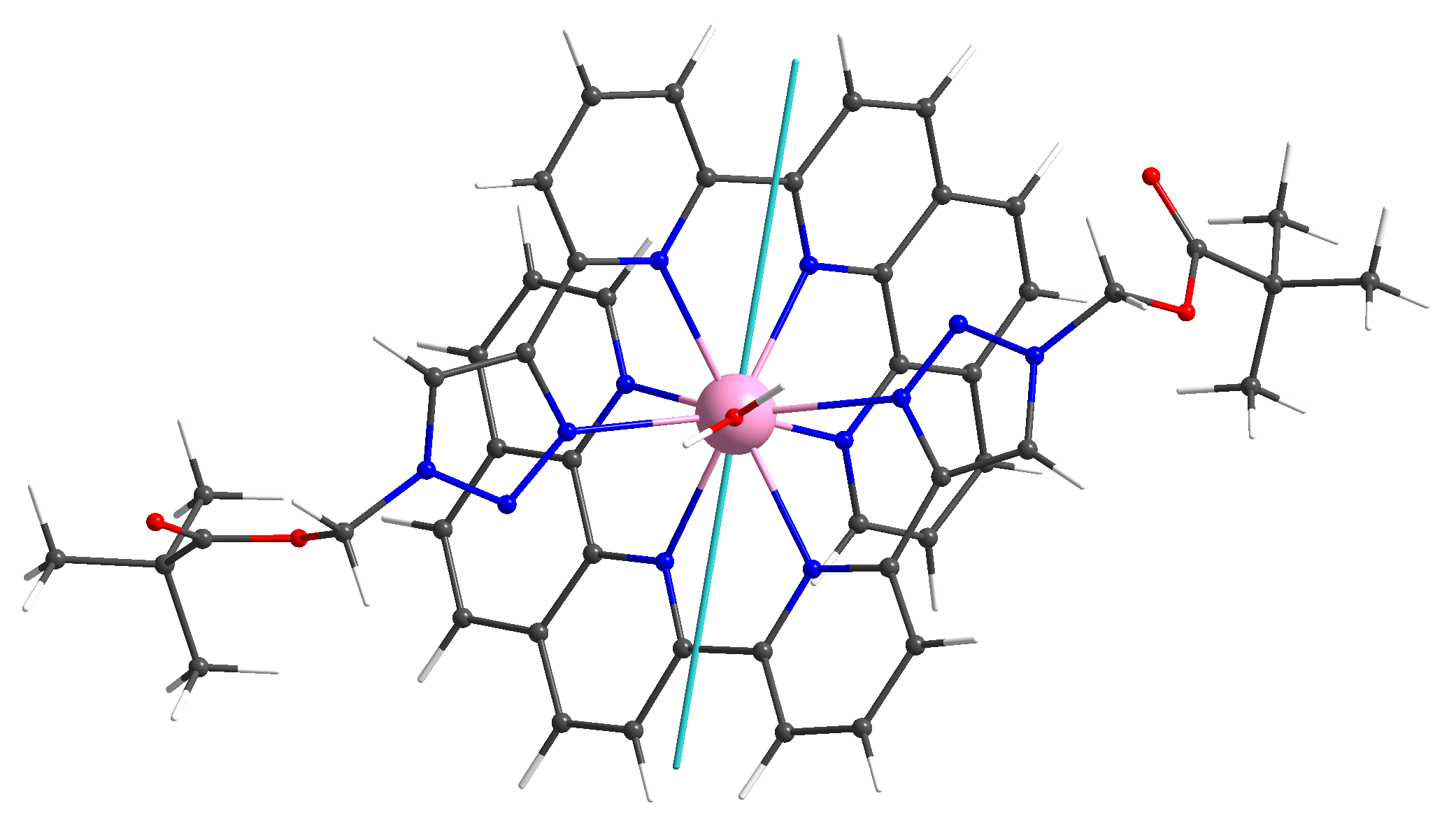}
    \caption{Orientation of the magnetic easy anisotropy axis, as indicated by the light blue axis. Counterions and methanol are not shown for simplicity, but included in the calculation.}
    \label{fig:anisotropy}
\end{figure}

The molecular magnetic easy anisotropy axis calculated for the full model is shown in figure \ref{fig:anisotropy}. It is orthogonal to the Er-O bond and is oriented through the two PPTMP ligands. It approximately penetrates the carbon-carbon bonds between the pyridine and phenanthroline moieties. Considering the prolate $4f$ electron distribution of \erbium , the main anisotropy axis points in the direction of the minimal electrostatic field~\cite{rinehart.2011}. In \one , the water molecule exerts the largest electrostatic field. Therefore, the main axis must be orthogonal to the Er-O bond. In this plane, the magnetic main axis will go along the minimal crystal field. This is corroborated by the molecular symmetry being close to $C_2$, which allows only for anisotropy axes parallel and perpendicular to the Er-O bond which acts as the rotational axis. The resulting $\chi T$-product and isothermal magnetisation are depicted in Fig.~\ref{fig:DCmag}. The anisotropy axes of models neglecting additional counterions and MeOH are perpendicular to the Er-O bond, too. However, their axes are oriented almost in parallel to the position of the water ligand's hydrogen positions, as depicted in the Supplementary Material.

For the first excited KD of the full model, a $g_z$-value of 9.2 is calculated. This value is not directly comparable to the $g_\mathrm{eff,z}$ obtained from the loose powder HF-EPR data, since in the experiment, the crystallites are oriented along the main magnetic anisotropy axis of the ground state. The calculated effective $g_\mathrm{eff,z}$, therefore, is given by the projection of the real $g_\mathrm{z}$-value onto the magnetic field direction. This leads to $g_\mathrm{eff,z} = 7.0$ for KD2. In comparison, the calculated and projected $g_z$-values for the bare molecules with and without optimising the hydrogen positions are 1.8 and 2.7, respectively, which are far off the experimentally determined values. However, the calculated and projected $g_z$-value for the full model, without optimising hydrogen positions, is 5.8, which is very close to the experimental value. A summary of the orientations of the main anisotropy axes for the ground state KD of selected models can be found in the Supplementary Material.

\begin{figure}[t]
    \centering
    \includegraphics[width = \columnwidth]{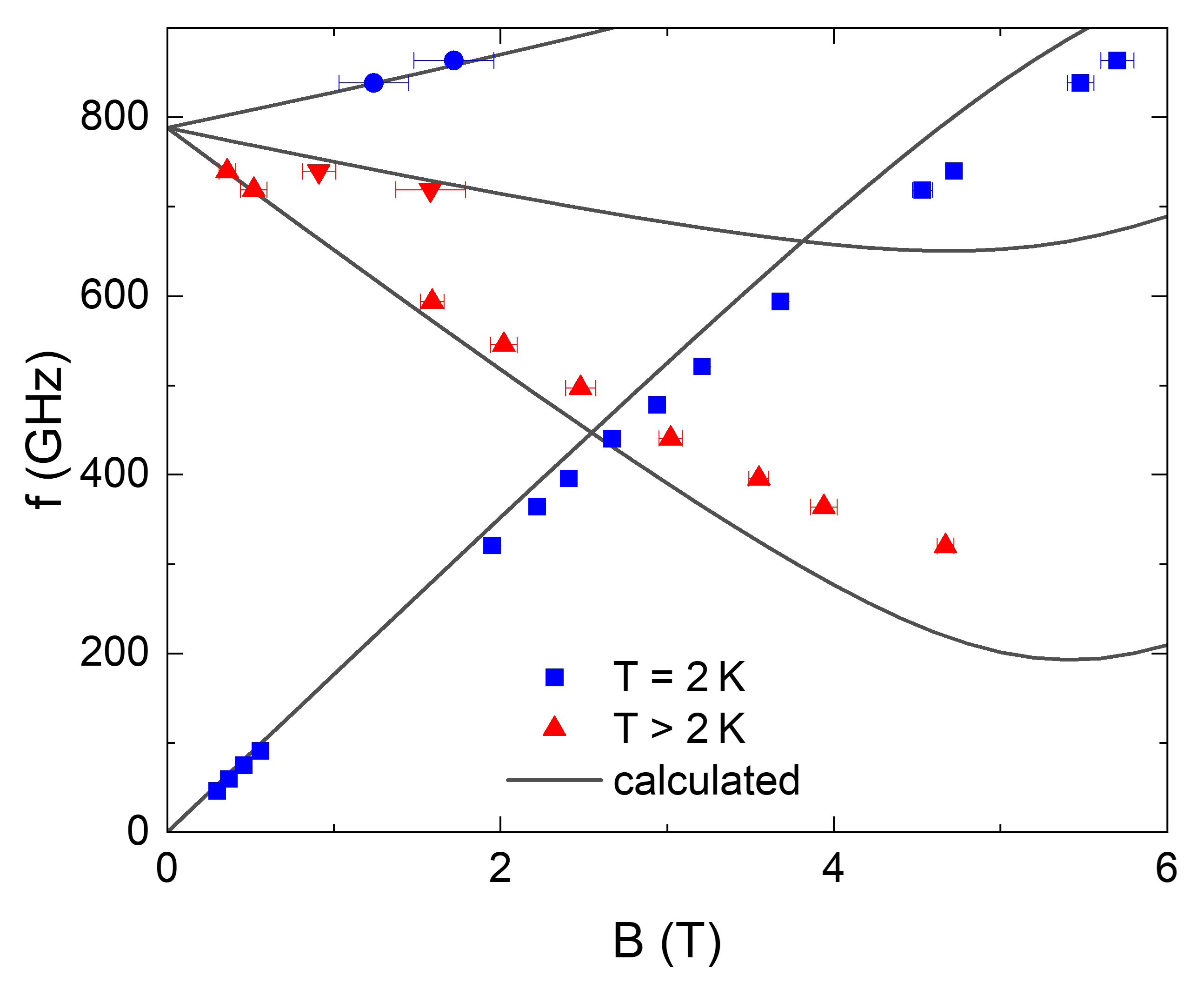}
    \caption{Measured (points) and calculated (lines) transition energies of a loose powder sample of \one\ within the lowest two Kramers doublets for the model including all counterions and optimised hydrogen positions.}
    \label{fig:zeemansim}
\end{figure}

By applying a magnetic field in the direction of the ground state magnetic easy anisotropy axis for the model with optimised hydrogen positions and all counterions, we are able to simulate the underlying Zeeman diagrams for the loose powder experiment as shown in the Supplementary Material. The transition energies between the four lowest states are shown in Fig.~\ref{fig:zeemansim}. We predict an avoided level crossing close to a field of 6~T with first deviations of the linear behaviour starting at around 4~T. This is in excellent agreement compared to the experimental measurements. The simulated versus experimentally observed Zeeman-splitting for the unoptimised model containing all counterions is shown in the Supplementary Material. Considering that the agreement is worse compared to Fig.~\ref{fig:zeemansim}, we will rule out the unoptimised model. 

The mixing of $m_J$-states is calculated for the eight Kramers Doublets. The ground state is mainly a mixture of the $m_J=\pm15/2\,(\approx31~\%)$ and $m_J=\pm9/2\,(\approx51~\%)$ states. In contrast, the first excited state dominantly consists of the $m_J=\pm7/2\,(\approx56~\%)$ and $m_J=\pm5/2\,(\approx22~\%)$ states. These values are in good agreement with the experiment but depend on the used model to a certain extent. For example, using the bare molecule and no optimisation of H-positions will lead to a ground state that is a mixture of $m_J$ = 15/2, 9/2 and 5/2 with 26~\%, 33~\%, and 20~\%, respectively, while the first excited state is an equal mixture of the $m_J$ = 7/2, 3/2 and 1/2 states with contributions between 24~\% and 28~\%. In summary, the comparison of the different models shows that the methanol as well as the three counterions should be considered to obtain the correct magnetic axes and g-values, while the energy differences are already well described by the bare molecule.

\section{Discussion and Conclusion}

A summary of the effective $g$-tensors and crystal field splittings as obtained by HF-EPR measurements and \textit{ab-initio} calculations is given in Tab.~\ref{tab:summary}. Our experimental and theoretical results clearly evidence a pronounced uniaxial behaviour, i.e.,~$g_\mathrm{eff,z} \gg g_\mathrm{eff,x},\,g_\mathrm{eff,y}$, of the energetically lowest Kramers doublet in \one\ and, hence, imply an Ising-like character of the \erbium\ moment in the magnetic ground state. According to the calculations, the water ligand is responsible for the largest contribution to the ligand field. Therefore, the direction of the Ising-axis is perpendicular to it. However, the observation of sizeable $g_\mathrm{eff,x/y}$ components in the effective $g$-tensor directly reveals a significant deviation from a perfect Ising scenario due to the presence of a considerable transversal crystal field. As described above, such off-diagonal elements in the single-ion anisotropy tensor lead to a mixing of the $m_J$ states and, thereby, give rise to avoided level crossings and fast magnetic relaxation via QTM in zero field. As demonstrated by the ac susceptibility data, the QTM in \one\ can be effectively quenched by shifting the energy levels of the initial and final state using a small magnetic field. In this in-field configuration, magnetic relaxation is dominated by temperature-activated processes, such as Orbach, Raman or thermally-assisted QTM. While KD3 is strongly gapped $\Delta_\mathrm{2 \rightarrow 3} \gg 1$~THz, the first excited Kramers doublet KD2 lies approximately $37$~K (780 GHz) above the ground state and can easily contribute to the observed slow magnetic relaxation.

\begin{table}[h]
    \centering
    \caption{Experimentally and theoretically obtained effective $g$-tensors and zero field splittings of the two lowest KDs in  \one .} 
    \begin{tabular}{cc|c|c}
          \hline \hline
          & & HF-EPR & CASOCI \\ \hline \hline
          & $g_\mathrm{eff,x}$ & 1.2(3) & 1.1 \\
          KD1 & $g_\mathrm{eff,y}$ & 2.6(4) & 2.0 \\
          & $g_\mathrm{eff,z}$ & 11.8(2) & 12.6 \\ \hline
          \multicolumn{2}{c|}{$\Delta_\mathrm{1 \rightarrow 2}$ (GHz)} & 780(5) & 788 \\ \hline
          KD2 & $g_\mathrm{eff,z}$ & 5.4(3) & 7.0 \\ \hline
          \multicolumn{2}{c|}{$\Delta_\mathrm{2 \rightarrow 3}$ (GHz)} & $\gg 1000$ & 2495 \\
          \hline \hline
    \end{tabular}
    \label{tab:summary}
\end{table}

The effective $g$-factors obtained from the pseudo-single-crystal loose powder spectra can be used to determine the effective magnetic quantum numbers $m_\mathrm{eff} = g_\mathrm{eff,z}/2g_\mathrm{L}$ for the experimentally accessible Kramers doublets~\cite{spillecke2021,arneth2025}. Assuming that the free ion Land\'{e} factor $g_\mathrm{L} = 1.2$ is still a good approximation for \erbium\ moments in the present coordination geometry, our analysis yields $m_\mathrm{eff,1} = 4.9(1)$ and $m_\mathrm{eff,2} = 2.3(1)$ for KD1 and KD2, respectively. The almost integer value of $m_\mathrm{eff,1}$ for KD1 confirms the occurrence of pronounced state mixing, as also evidenced by the observation of avoided level crossing in the frequency-field diagram. Furthermore, $m_\mathrm{eff,1}$ is close to 4.5, which indicates that the low-symmetry crystal field in \one\ stabilises the $m_J = \pm 9/2$ state as the magnetic ground state. As for KD2, the main contribution to the spin wavefunction is suggested to be $m_J = 5/2$.

As already evident from the good agreement of the measured and calculated dc magnetisation, the experimentally determined $g$-tensor for KD1 is well reproduced by CASOCI. While CASOCI also gives an accurate prediction of the zero-field gap to the first excited Kramers doublet, the effective $g_\mathrm{z}$-value for KD2 is slightly overestimated by this approach. In agreement with the absence of corresponding transitions in the HF-EPR experiments, the calculation finds KD3 to be strongly gapped with $\Delta_\mathrm{2 \rightarrow 3} \simeq 2.5$~THz, i.e., well beyond the experimentally accessible range.

To sum up, we successfully synthesised and magnetically characterised a triply-capped, slightly distorted trigonal-prismatic \erbium\ complex \one\ using combined dc and ac magnetic, HF-EPR and numerical studies. Complex \one\ is SMM-silent due to QTM in zero field, but exhibits field-induced slow relaxation of the magnetisation at $B = 0.1$~T. The dominance of under-barrier relaxation is rationalised by the presence of sizeable transversal components in the overall uniaxial effective $g$-tensor of the ground state KD1. When QTM is quenched, slow magnetic relaxation happens via thermally activated processes involving the first excited KD2 at $780$~GHz above the ground state. KD3 is shown to be strongly gapped and, thus, does not contribute to the relaxation dynamics at low temperatures. In particular, our detailed spectroscopic investigations allow for a direct experimental determination of the effective $g$-tensors and zero-field splittings of the two energetically lowest Kramers doublets. The excellent agreement between the measured data and the results of the CASOCI calculations for g-tensors and zero-field splitting support the reliability of both approaches.

\section{Experimental Details and Methods}
Polycrystalline powder samples of \one\ have been synthesised as described in the Supplemental Information.

For crystal structure determination a suitable crystal of \one\ was covered in mineral oil (Aldrich) and mounted on a glass fibre. The crystal was transferred directly to the cold stream of a STOE IPDS 2 ($150$~K) or STOE StadiVari ($100$ or $110$~K) diffractometer. All structures were solved by using the program SHELXS/T~\cite{sheldrick2007,sheldrick2015} and Olex~\cite{dolomanov2009}. The remaining non-hydrogen atoms were located from successive difference Fourier map calculations. The refinements were carried out by using full-matrix least-squares techniques on $F^2$ by using the program SHELXL~\cite{sheldrick2007,sheldrick2015}. The H-atoms were introduced into the geometrically calculated positions (SHELXL procedures) unless otherwise stated and refined riding on the corresponding parent atoms. In each case, the locations of the largest peaks in the final difference Fourier map calculations, as well as the magnitude of the residual electron densities, were of no chemical significance. Summary of the crystal data, data collection and refinement for compounds are given in the supplement. Crystallographic data for the structures reported in this paper halve been deposited with the Cambridge Crystallographic Data Centre as a supplementary publication no.~CCDC 2472872. Copies of the data can be obtained free of charge on application to CCDC, 12 Union Road, Cambridge CB21EZ, UK (fax: (+(44)1223-336-033; email: deposit@ccdc.cam.ac.uk).

The direct current (dc) and alternating current (ac) magnetisation was studied in the temperature range $T = 1.8-300$~K by means of a Magnetic Properties Measurement System (MPMS3, Quantum Design) and a Physical Properties Measurement System (PPMS, Quantum Design) in magnetic fields up to 7~T and 14~T, respectively. For all measurements powder samples were pelletised in polycarbonate capsules to avoid reorientation in external magnetic fields. The experimental data were corrected for the contribution of the sample holder and of the diamagnetic ligand field calculated by means of Pascal's constants~\cite{bain2008}. Simulations of the magnetic data were performed using the PHI software package~\cite{phi}.

Multifrequency high-field electron paramagnetic resonance studies were performed using a phase-sensitive millimeter vector network analyser (MVNA) by ABmm as a simultaneous microwave source and detector~\cite{comba2015}. Temperature control from $2$~K to $300$~K was ensured by placing the sample space in the Variable Temperature Insert (VTI) of an Oxford magnet system equipped with a $16$~T superconducting coil~\cite{werner2017}. Polycrystalline powder samples were prepared in a brass ring sealed with kapton tape either as loose powder or fixed by mixing with eicosane. The former setup allows alignment of the crystallites with the external magnetic field, hence providing simplified pseudo-single-crystal spectra (see Refs.~\cite{goldberg1997,barra1996,spillecke2021,ahmed2022,arneth2025,butterflies2025}). Alignment is ensured by sweeping up the magnetic field to $16$~T prior to each measurement and confirmed by observation of corresponding orientation effects in the transmitted microwave intensity signal. Spectral simulations of the HFEPR data were performed using the EasySpin sofware package~\cite{stoll2006}.

\textit{Ab-initio} Complete Active Space Spin-orbit Configuration Interaction (CASOCI) calculations~\cite{bodenstein.2022} are carried out in the following steps: First, the hydrogen positions of the experimentally determined structure are optimized on a B3-LYP-d4/def2-TZVP~\cite{becke.1993,caldeweyher.2017,caldeweyher.2019} level of theory in TURBOMOLE~\cite{balasubramani.2020,tm.2007}, by replacing Er to Y. For each model complex investigated, Y is replaced to Gd and an unrestricted Kohn Sham calculation is performed using the B3-LYP functional, the all-electron basis x2c-TZVPall for Gd, the def2-TZVP basis for the remaining atoms and an energy convergence criterium of $10^{-7}\ E_h$. Scalar relativistic effects are accounted for using X2C~\cite{peng.2013}. The half filled natural orbitals of this calculation are used as a starting point for a Restricted Open Shell Hartree Fock (ROHF) calculation on the same compound, now using the second order Douglass Kroll Hess approach to treat scalar relativstic effects~\cite{vanwullen.2004,vanwullen.2005}. This ensures a reliable starting guess for pure $f$ orbitals in the active space. In a next step, Gd is replaced to Er and the orbitals of the previous calculation are used as a starting guess. The Roothaan parameters used are $a=119/121$ and $b=126/121$ mimicking an average over all quartet states with eleven electrons in the seven 4$f$ orbitals for the orbital optimization. In a last step DKH2 is changed to DKH4 and a density convergence criterium of $10^{-7}$ is introduced keeping the energy convergence criterium of $10^{-7}\ E_h$ for a tight convergence of the calculation. We have tested the influence of a subsequent state averaged CASSCF calculation for optimizing the orbitals explicitly for the $^4F$-ground term. Afterwards the CASOCI calculations were carried out explicitly introducing spin-orbit coupling. This is done by diagonalizing the spin-orbit coupled Full CI matrix in an active space of eleven electrons in the seven 4$f$ orbitals (11,7) active space using a Davidson algorithm to extract the lowest 100 eigenstates~\cite{bodenstein.2022}.

{\bf Supporting Information:} The Supporting Information contains additional details about the synthesis; structure figures and X-ray crystallographic data; NMR, IR, mass spectroscopy data; it also includes Ref.~[\citenum{zong2005}]. Further information about the numerical results (file: ErIII-SI.pdf). In addition, it includes xyz-files on all model structures used in this manuscript. The difference between optimized and unoptimized is that in
optimized structures, hydrogen positions were optimized on a DFT level as described in the text. This was done considering all
counterions and lattice solvent, so on the "full" structure. Afterwards, the model was truncated to generate the other
structures (file: ErIII-SI-xyz.zip).


\begin{acknowledgements}
We thank Florian Bruder and Florian Weigend for valuable discussions and scientific input. David Frick is acknowledged for solving the single crystal X-ray structure of compound \one . Support by Deutsche Forschungsgemeinschaft (DFG) under Germany's Excellence Strategy EXC2181/1-390900948 (the Heidelberg STRUCTURES Excellence Cluster) is gratefully acknowledged. J.~A.~acknowledges support by the International Max-Planck Research School for Quantum Dynamics (IMPRS-QD) Heidelberg. Financial support for this research was provided by the German Federal Ministry of Education and Research through the f-Char project under grant numbers 02NUK059A. C.~P.~acknowledges the state of Baden-Württemberg via the Landesgraduiertenförderung and the DFG via the CRC 1573 "4f for future" for funding.
\end{acknowledgements}

\bibliography{Er_muffin}

\clearpage
\newpage

\onecolumngrid

\section*{Supplemental Information}\label{Sec_Appendix}


\title{Supplemental Information:\\ Towards Understanding Prolate 4$f$ Monomers: Numerical Predictions and Experimental Validation of Electronic Properties and Slow Relaxation in a Muffin-shaped Er$^\mathrm{III}$ Complex}

\author{J.~Arneth$^+$}\email{jan.arneth@kip.uni-heidelberg.de}
\affiliation{Kirchhoff Institute for Physics, Heidelberg University, INF 227, D-69120 Heidelberg, Germany}
\author{C. Pachl$^+$}
\affiliation{Institute of Nanotechnology, Karlsruhe Institute of Technology (KIT),  Kaiserstr. 12, 76131 Karlsruhe, Germany}
\affiliation{Institute of Inorganic Chemistry, Karlsruhe Institute of Technology,  Kaiserstr. 12, 76131 Karlsruhe, Germany}
\author{G. Greif}
\affiliation{Institute of Inorganic Chemistry, Karlsruhe Institute of Technology, Kaiserstr. 12, 76131 Karlsruhe, Germany}
\author{B.~Beier}
\affiliation{Kirchhoff Institute for Physics, Heidelberg University, INF 227, D-69120 Heidelberg, Germany}
\author{P. W. Roesky}\email{roesky@kit.edu}
\affiliation{Institute of Inorganic Chemistry, Karlsruhe Institute of Technology,  Kaiserstr. 12, 76131 Karlsruhe, Germany}
\affiliation{Institute of Nanotechnology, Karlsruhe Institute of Technology (KIT),  Kaiserstr. 12, 76131 Karlsruhe, Germany}
\author{K. Fink}\email{karin.fink@kit.edu}
\affiliation{Institute of Nanotechnology, Karlsruhe Institute of Technology (KIT),  Kaiserstr. 12, 76131 Karlsruhe, Germany}
\author{R.~Klingeler}\email{klingeler@kip.uni-heidelberg.de}
\affiliation{Kirchhoff Institute for Physics, Heidelberg University, INF 227, D-69120 Heidelberg, Germany}

\maketitle
\def\thefootnote{+}\footnotetext{These authors contributed equally to this work.}\def\thefootnote{\arabic{footnote}}

\renewcommand{\thefigure}{S\arabic{figure}}
\setcounter{figure}{0}
\setcounter{section}{0}

The Supporting Information contains additional details about the synthesis, structure figures and X-ray crystallographic data as well as NMR, IR, mass spectroscopy data. Further information about the numerical results can be found in an additional Supplemental File (ErIII-SI-xyz.zip). The latter includes xyz-files on all model structures used in this paper. The difference between optimized and unoptimized is that in
optimized structures, hydrogen positions were optimized on a DFT level as described in the paper. This was done considering all
counterions and lattice solvent, so on the "full" structure. Afterwards, the model was truncated to generate the other
structures.

\section{Synthesis and Crystal Structure}

All chemicals were obtained from commercial sources and were used without further purification. \\
2-(6-((Trimethylsilyl)ethynyl)pyridin-2-yl)-1,10-phenanthroline was prepared according to a literature procedure~\cite{zong2005}.

\subsection{(4-(6-(1,10-phenanthrolin-2-yl)pyridin-2-yl)-1H-1,2,3-triazol-1-yl)methyl pivalate (PPTMP)}

\begin{figure}[h]
    \centering
    \includegraphics[width=0.7\columnwidth]{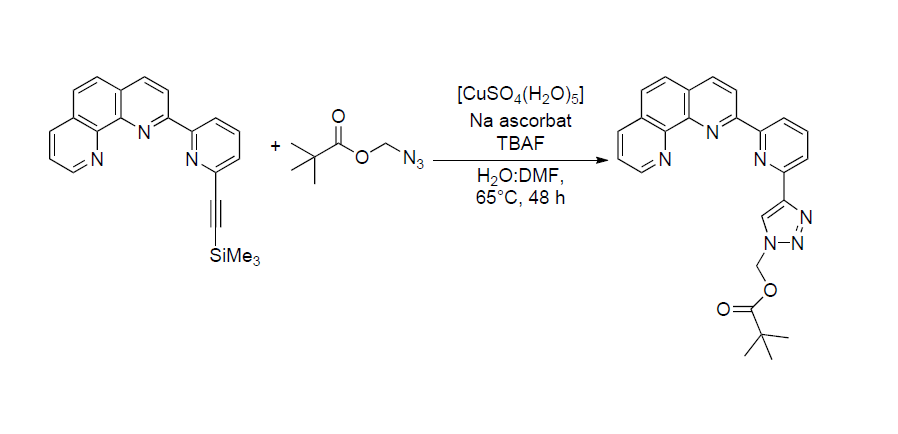}
    \caption{Synthesis scheme of PPTMP.}
    \label{fig:synthesisPPTMP}
\end{figure}

2-(6-((Trimethylsilyl)ethynyl)pyridin-2-yl)-1,10-phenanthroline (899~mg, 2.54~mmol, 1.00~eq.), azidomethyl pivalate (479~mg, 2,80~mmol, 1.10~eq.), sodium ascorbate (65.5~mg, 0.331~mmol, 0.13~eq.) and copper(II) sulfate pentahydrate (25.4~mg, 0.102~mmol, 0.04~eq.) were dissolved in a DMF:H$_2$O-mixture (11~mL; ratio 4:1) and dropwise treated with tetrabutylammonium fluoride (2.54~mL, 665~mg, 1~M in THF, 2.54~mmol, 1.00~eq.). The reaction mixture was then stirred at 65~°C for 16~h. After cooling the mixture to room temperature, brine (200~mL) was added and the solution was extracted with ethyl acetate (2×200~mL). The combined organic phases were washed with brine (2×200~mL) and dried over Na$_2$SO$_4$. The solvent was removed under reduced pressure to give a brown oil which was purified by column chromatography (hexane: ethyl acetate). The product was obtained as a brown solid (718~mg, 1.64~mmol, 65~\%).

{\bf ${}^1$H-NMR (400~MHz, CDCl$_3$):} $\delta$ [ppm] = 9.25 (dd, ${}^3J_{HH}$ = 4.3, 1.8 Hz, 1H, CH$_\mathrm{Php}$), 8.98 (dd, ${}^3J_{HH}$ = 7.9, 1.1 Hz, 1H, CH$_\mathrm{Py}$), 8.90 (d, ${}^3J_{HH}$ = 8.4 Hz, 1H, CH$_\mathrm{Py}$), 8.55(s, 1H, CH$_\mathrm{Tri}$), 8.38 (d, ${}^3J_{HH}$ = 8.4 Hz, 1H, CH$_\mathrm{Php}$), 8.26 (ddd, ${}^3J_{HH}$ = 7.8, 2.9, 1.5 Hz, 2H, CH$_\mathrm{Php}$), 8.01 (t, ${}^3J_{HH}$ = 7.8 Hz, 1H, CH$_\mathrm{Php}$), 7.82 (m, 2H, CH$_\mathrm{Php}$), 7.65 (dd , ${}^3J_{HH}$ = 8.0, 4.3 Hz, 1H, CH$_\mathrm{Py}$), 6.35 (s, 2H, CH$_\mathrm{2}$), 1.22 (s, 9H, tBu).

{\bf ${}^{13}$C{${}^1$H} NMR (101 MHZ, CDCl$_3$):} $\delta$ [ppm] = 178.0 (C$_\mathrm{q}$), 156.0 (C-C$_\mathrm{q}$), 155.9 (C$_\mathrm{q}$), 150.6 (CH$_\mathrm{Php}$), 149.3 (C$_\mathrm{q}$), 149.2 (C$_\mathrm{q}$), 146.5 (C$_\mathrm{q}$), 145.8 (C$_\mathrm{q}$), 138.1 (CH$_\mathrm{Php}$), 137.0 (CH$_\mathrm{Php}$), 136.4 (CH$_\mathrm{Php}$), 129.2 (C$_\mathrm{q}$), 129-0 (C$_\mathrm{q}$), 127.0 (CH$_\mathrm{Php}$), 126.7 (CH$_\mathrm{Php}$), 123.7 (CH$_\mathrm{Tri}$), 123.1 (CH$_\mathrm{Py}$), 122.3 (CH$_\mathrm{Py}$), 121.3 (CH$_\mathrm{Py}$), 120.9 (CH$_\mathrm{Php}$), 69.9 (CH$_\mathrm{2}$), 39.0 (C$_\mathrm{q}$), 27.0 (CH$_\mathrm{3}$).

{\bf IR (ATR):} $\bar{\nu}$ [cm$^{-1}$] = 3112 (vw), 2974 (vw), 2932 (vw), 2871 (vw), 1745 (w), 1600 (vw), 1557 (vw), 1505 (vw), 1492 (vw), 1457 (vw), 1439 (vw), 1411 (w), 1367 (vw), 1276 (w), 1256 (w), 1187 (vw), 1121 (m), 1061 (vw), 1033 (w), 1015 (w), 991 (w), 943 (vw), 884 (vw), 867 (w), 828 (w), 796 (w), 772 (w), 747 (w), 720 (w), 681 (vw), 657 (vw), 645 (vw), 623 (vw), 579 (vw), 531 (vw), 509 (vw), 484 (vw), 448 (vw), 418 (vw). 

{\bf ESI-MS [m/z]} (C$_{25}$H$_{23}$N$_6$O$_2$, [M+H]$^+$): calc. 439.18, found 439.18. 

\subsection{[Er(PPTMP)$_2$(H$_2$O)][OTf]$_3$}

\begin{figure}[h]
    \centering
    \includegraphics[width=0.7\columnwidth]{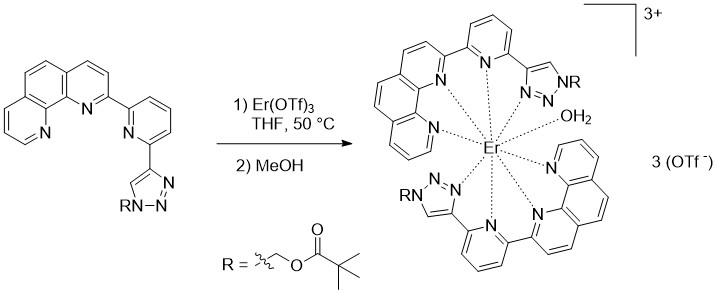}
    \caption{Synthesis scheme of complex \one .}
    \label{fig:synthesis_complex_supp}
\end{figure}

PPTMP (100 mg, 0.228 mmol, 1.00 eq.) and erbium(III) trifluoromethanesulfonate (140 mg, 0.288 mmol, 1.00 eq.) were dissolved in THF (10 mL) and heated overnight at 50 °C resulting in the formation of a pink precipitate. The supernatant was decanted and the precipitate was dried under vacuum. The precipitate was dissolved in hot methanol and the solution was allowed to evaporate slowly at room temperature. After a few days, pink crystals of the product were obtained. Yield: 53 mg, 34.4 mmol, 15~\%

{\bf IR (ATR):} $\bar{\nu}$ [cm$^{-1}$] = 3392 (m), 3257 (w), 2924 (vw), 2853 (vw), 1743 (vw), 1659 (w), 1642 (w), 1627 (w), 1576 (vw), 1523 (vw), 1499 (vw), 1476 (vw), 1431 (vw), 1391 (vw), 1223 (s), 1180 (m), 1082 (vw), 1025 (s), 859 (vw), 829 (vw), 808 (vw), 770 (vw), 746 (vw), 669 (vw), 631 (m), 577 (vw), 517 (w).

{\bf ESI-MS [m/z]} (C$_{52}$H$_{44}$ErF$_6$N$_{12}$O$_{10}$S$_2$, [M–OTf-H$_2$O]$^+$): calc. 1342.20, found. 1342.19.

\newpage

\section{Crystal Structure}

\begin{figure}[h]
    \centering
    \includegraphics[width=0.95\columnwidth]{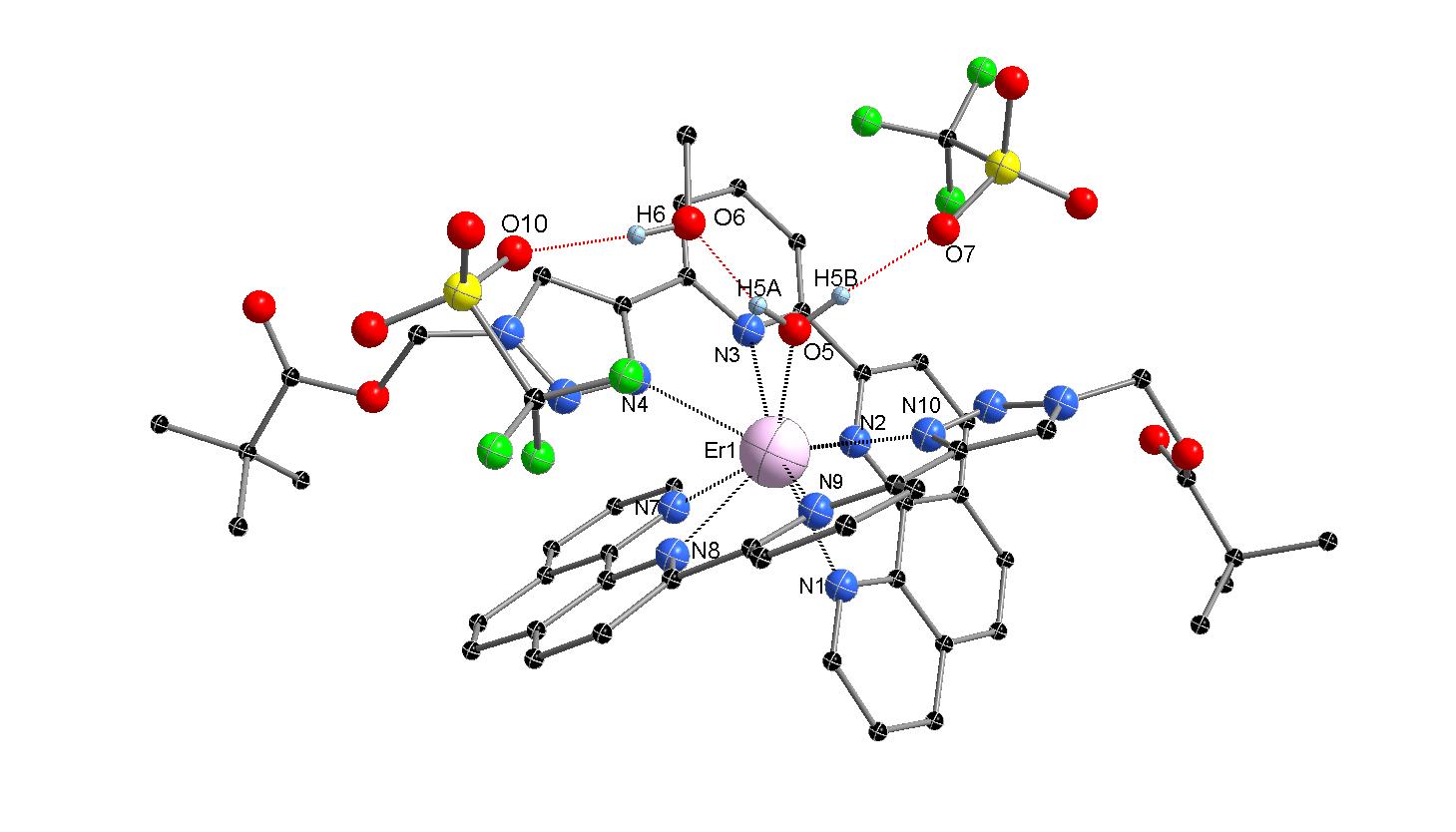}
    \caption{Molecular structures of [Er(PPTMP)2(H2O)][OTf]3 in the solid state. Thermal ellipsoids are drawn at 50~\% probability. Only selected hydrogen atoms are shown for clarity. Selected bond lengths (\AA) and angles ($^\circ$): Er1-N1 2.507(2), Er1-N2 2.473(2), Er1-N3 2.508(2), Er1-N4 2.529(2), Er1-N7 2.481(2), Er1-N8 2.473(2), Er1-N9 2.510(2), Er1-N10 2.4956(19), Er1-O5 2.3550(16), H5A-O6 1.8151(21), H5B-O7 1.8596(21), H6-O10 1.9191(23), N1-Er1-N2 66.54(7), N2-Er1-N3 63.67(7), N3-Er1-N4 65.39(7), N7-Er1-N8 66.68(7), N8-Er1-N9 63.45(7), N9-Er1-N10 65.43(7).}
    \label{fig:struct2}
\end{figure}

\begin{table}[h]
    \centering
    \caption{Crystal data, data collection and refinement of compound [Er(PPTMP)$_2$(H$_2$O)][OTf]$_3$.}
    \begin{tabular}{cc}
        Compounds & [Er(PPTMP)$_2$(H$_2$O)][OTf]$_3$ \\
        Chemical formula & C$_{54}$H$_{50}$ErF$_9$N$_{12}$O$_{15}$S$_3$ \\
        CCDC Number & \\ 2472872 \\
        Formula Mass & 1541.50 \\
        Radiation type & MoK$_\alpha$ \\
        Wavelength/nm & 0.71073 \\
        Crystal system & triclinic \\
        a/\AA & 12.1106(4) \\
        b/\AA & 13.8842(5) \\
        c/\AA & 18.2987(7) \\
        $\alpha$/° & 83.778(3) \\
        $\beta$/° & 82.542(3) \\
        $\gamma$/° & 84.534(3) \\
        Unit cell volume/\AA$^3$ & 3021.07(2) \\
        Temperature/K & 100 \\
        Space group & $P\bar{1}$ \\
        Z & 2 \\
        Absorption coefficient, $\mu$/mm & 1.598 \\
        No.~of reflections measured & 40840 \\
        No.~of independent reflections & 15900 \\
        $R_\mathrm{int}$ & 0.0303 \\
        Final $R_1$ values ($I>2\sigma(I)$) & 0.0324 \\
        Final $wR(F^2)$ values ($I>2\sigma(I)$) & 0.0722 \\
        Final $R_1$ values (all data) & 0.0442 \\
        Final $wR(F^2)$ values (all data) & 0.0756 \\
        Goodness of fit on $F^2$ & 0.985
    \end{tabular}
    \label{tab:structData}
\end{table}

\clearpage

\section{NMR Spectroscopy}

NMR spectra were recorded on Bruker spectrometers (Avance Neo 300 MHz, Avance Neo 400 MHz or Avance III 400 mHz). All NMR spectra were measured at 298 K, unless otherwise specified. The multiplicity of the signals is indicated as s = singlet, d = doublet, dd = doublet of doublets, t = triplet, q = quartet, m = multiplet and br = broad.

\begin{figure}[h]
    \centering
    \includegraphics[width=0.7\columnwidth]{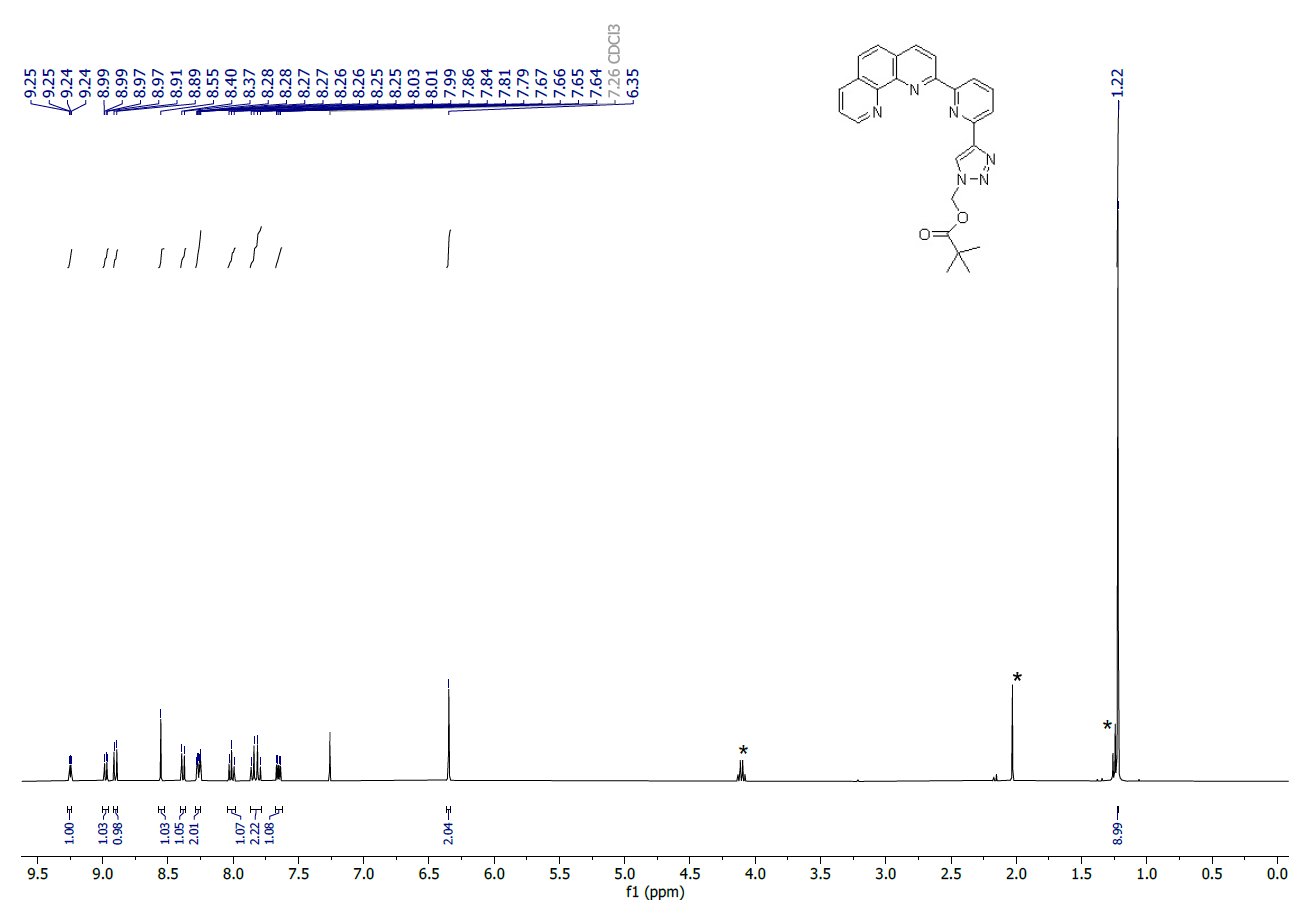}
    \caption{$^1$H NMR spectrum (400 mHz, CDCl$_3$, 298 K) of (4-(6-(1,10-phenanthrolin-2-yl)pyridin-2-yl)-1H-1,2,3-triazol-1-yl)methyl pivalate (PPTMP). *Solvent signal.}
    \label{fig:NMRPPTMP}
\end{figure}

\begin{figure}[h]
    \centering
    \includegraphics[width=0.7\columnwidth]{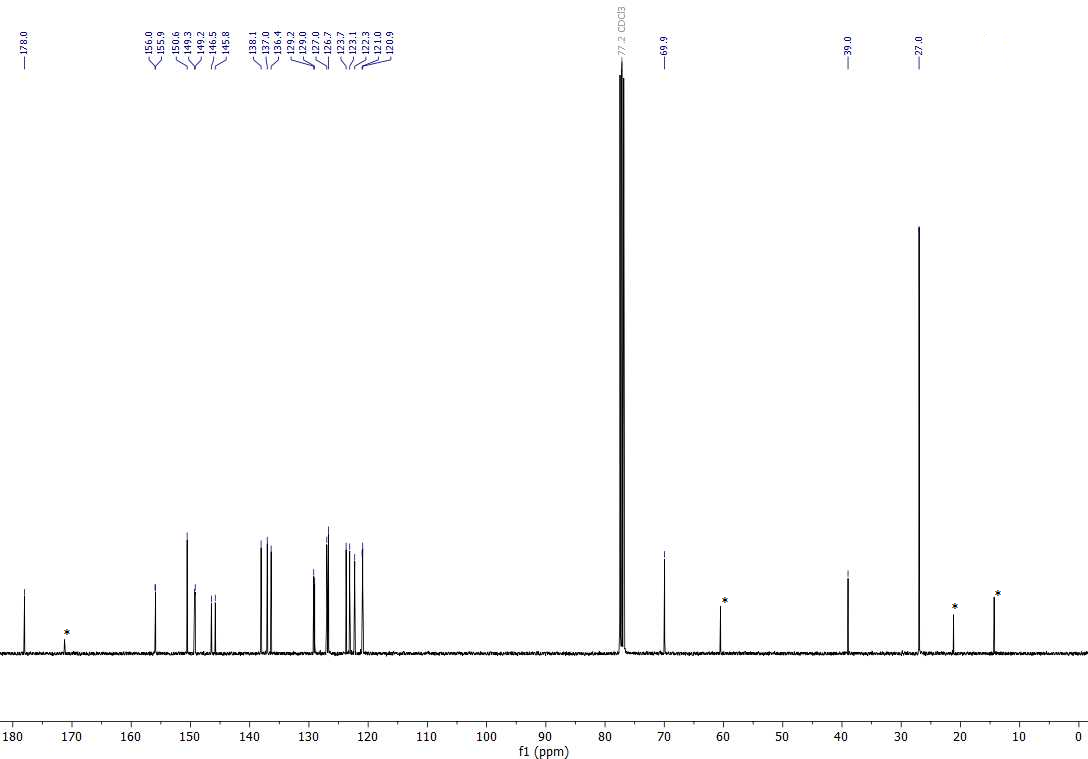}
    \caption{$^{13}$C{$^1$H} NMR spectrum (400 mHz, CDCl$_3$, 298 K) of PPTMP. *Solvent signal. }
    \label{fig:NMRC13}
\end{figure}

\clearpage

\section{IR Spectroscopy}

Infrared (IR) spectra were recorded in the region 4000–400 cm-1 on a Bruker Tensor 37 FTIR spectrometer equipped with a room temperature DLaTGS detector, a diamond attenuated total reflection (ATR) unit and a nitrogen-flushed chamber. In terms of their intensity, the signals were classified into different categories (vs = very strong, s = strong, m = medium, w = weak, and sh = shoulder).

\begin{figure}[h]
    \centering
    \includegraphics[width=0.7\columnwidth]{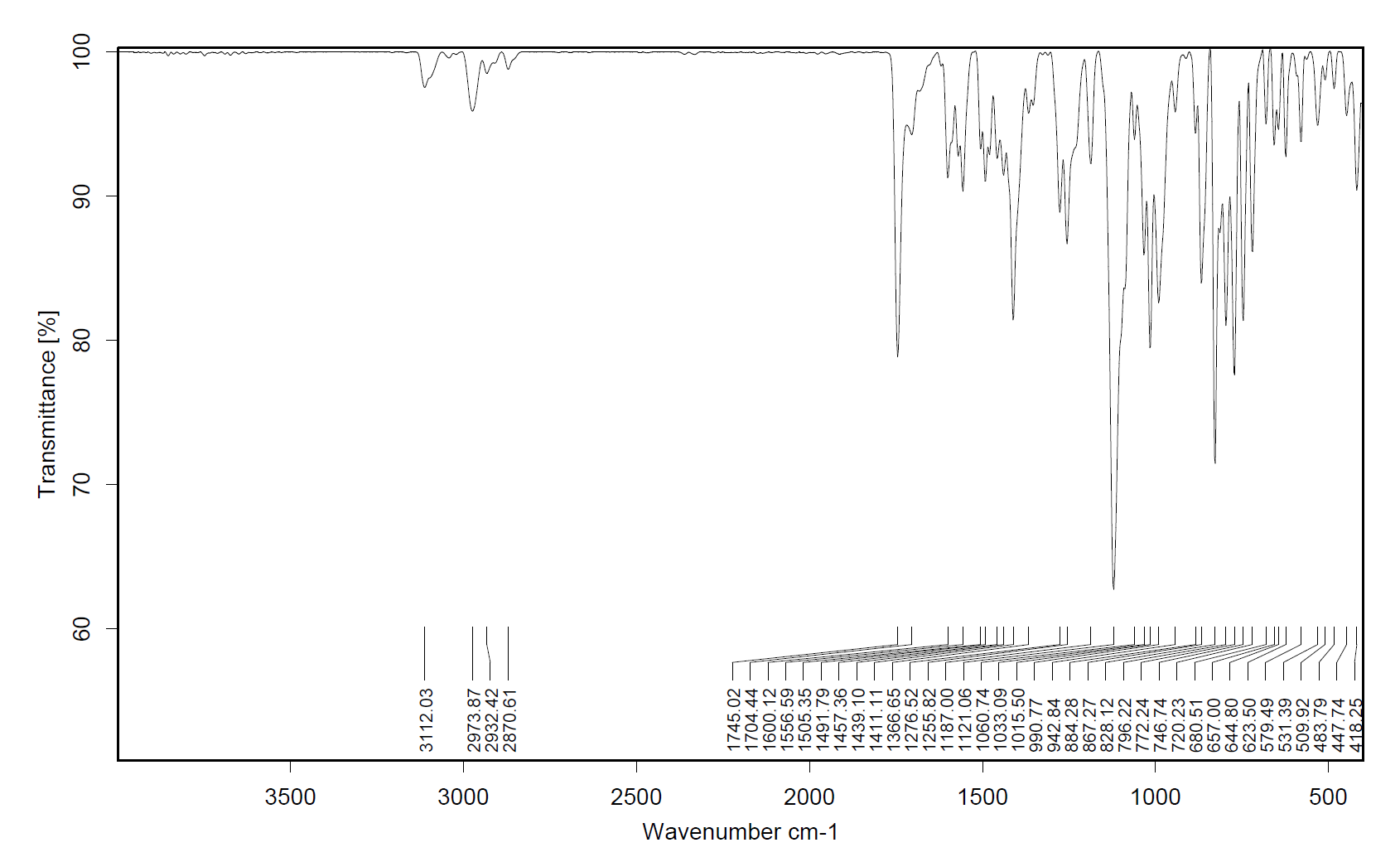}
    \caption{IR spectrum of PPTMP}
    \label{fig:IRPPTMP}
\end{figure}

\begin{figure}[h]
    \centering
    \includegraphics[width=0.7\columnwidth]{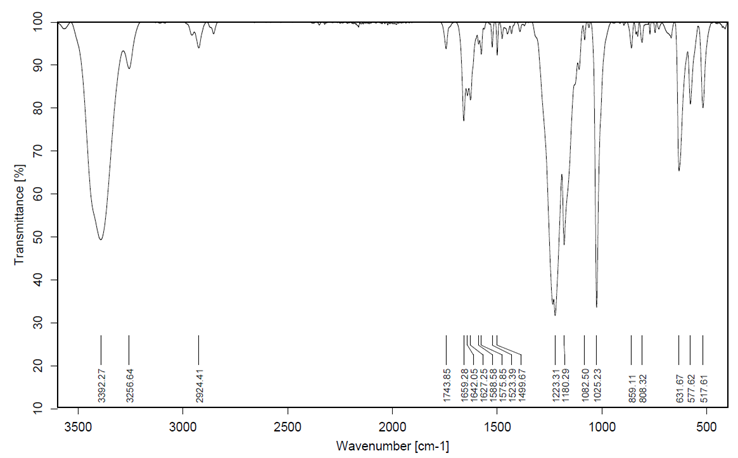}
    \caption{IR spectrum of [Er(PPTMP)$_2$(H$_2$O)][OTf]$_3$.}
    \label{fig:IRComplex}
\end{figure}

\clearpage

\section{Mass Spectrometry}

MS spectra were obtained on a Thermo Fisher Scientific LTQ Orbitrap XLQ Exactive mass spectrometer.

\begin{figure}[h]
    \centering
    \includegraphics[width=0.5\columnwidth]{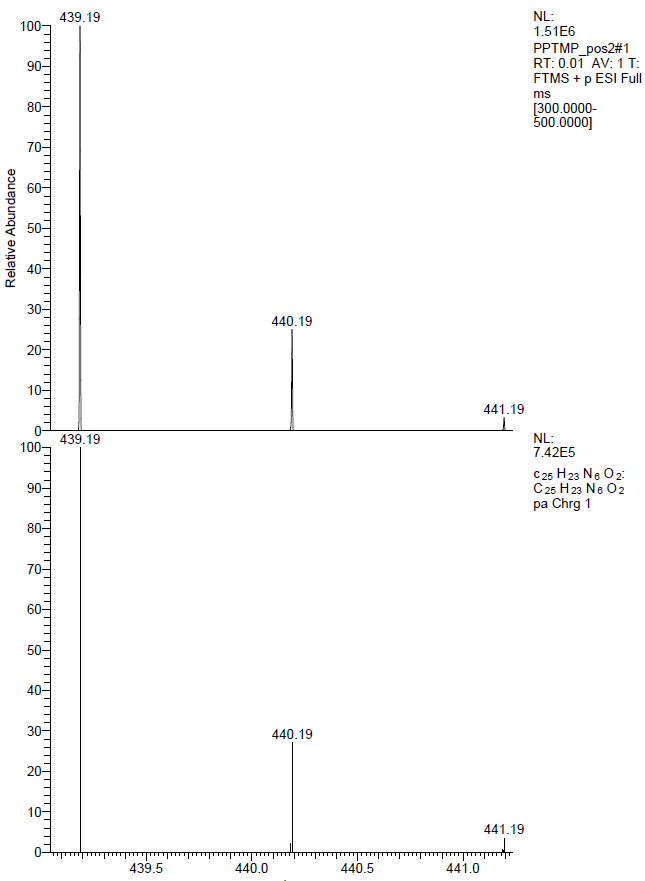}
    \caption{MS-ESI spectra of PPTMP. Shown is the molecular ion Peak [M+H]$^+$. Top: experimental spectrum, Bottom: simulated signals.}
    \label{fig:MSPPTMP}
\end{figure}

\begin{figure}[h]
    \centering
    \includegraphics[width=0.7\columnwidth]{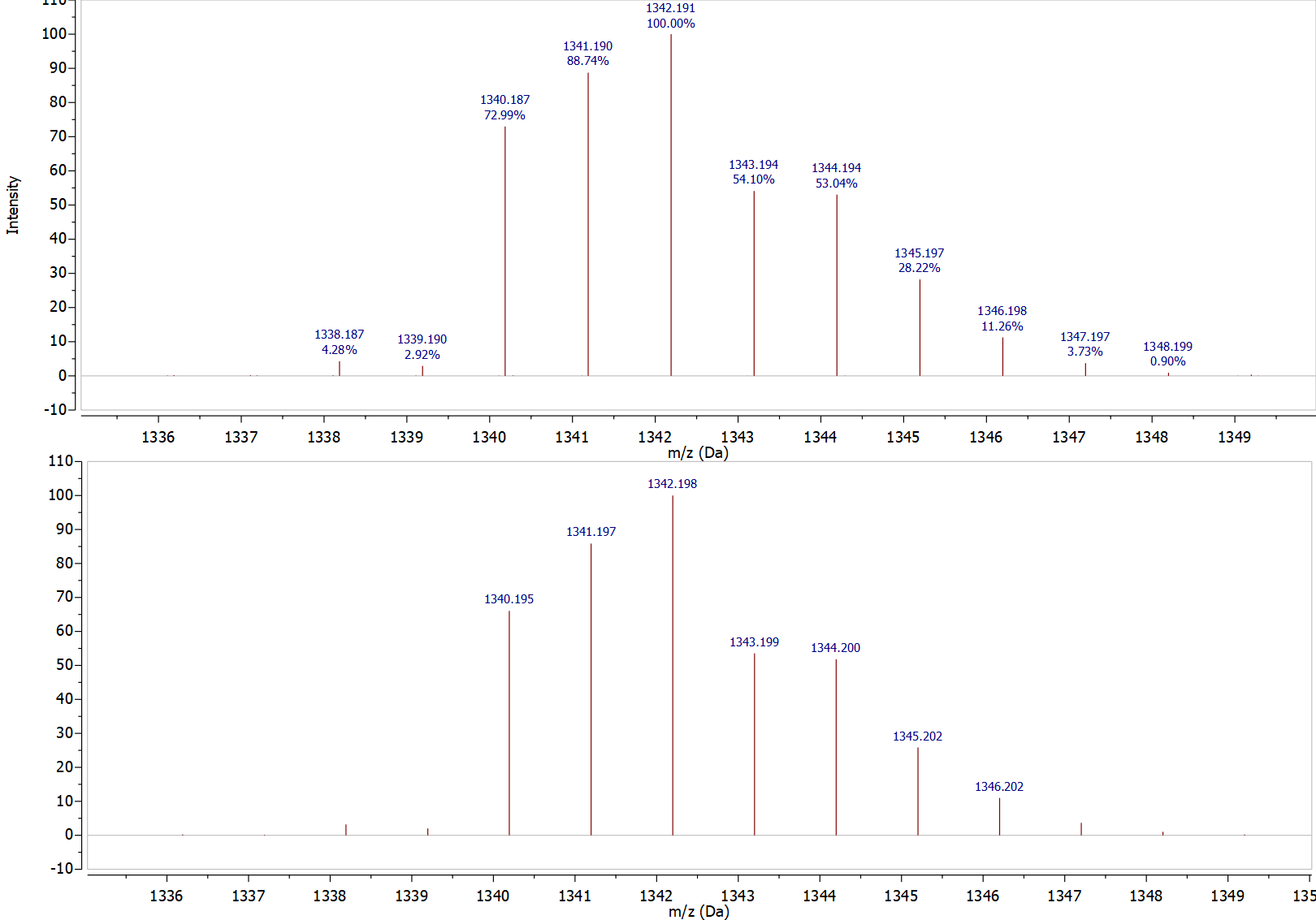}
    \caption{MS-ESI spectra of [Er(PPTMP)$_2$(H$_2$O)][OTf]$_3$. Shown is the molecular ion Peak [M – OTf-H$_2$O]$^+$. Top: experimental spectrum, Bottom: simulated signals.}
    \label{fig:MSComplex}
\end{figure}

\clearpage

\section{Calculations}

\begin{figure}[htb]
    \centering
    \includegraphics[height=0.77\textheight]{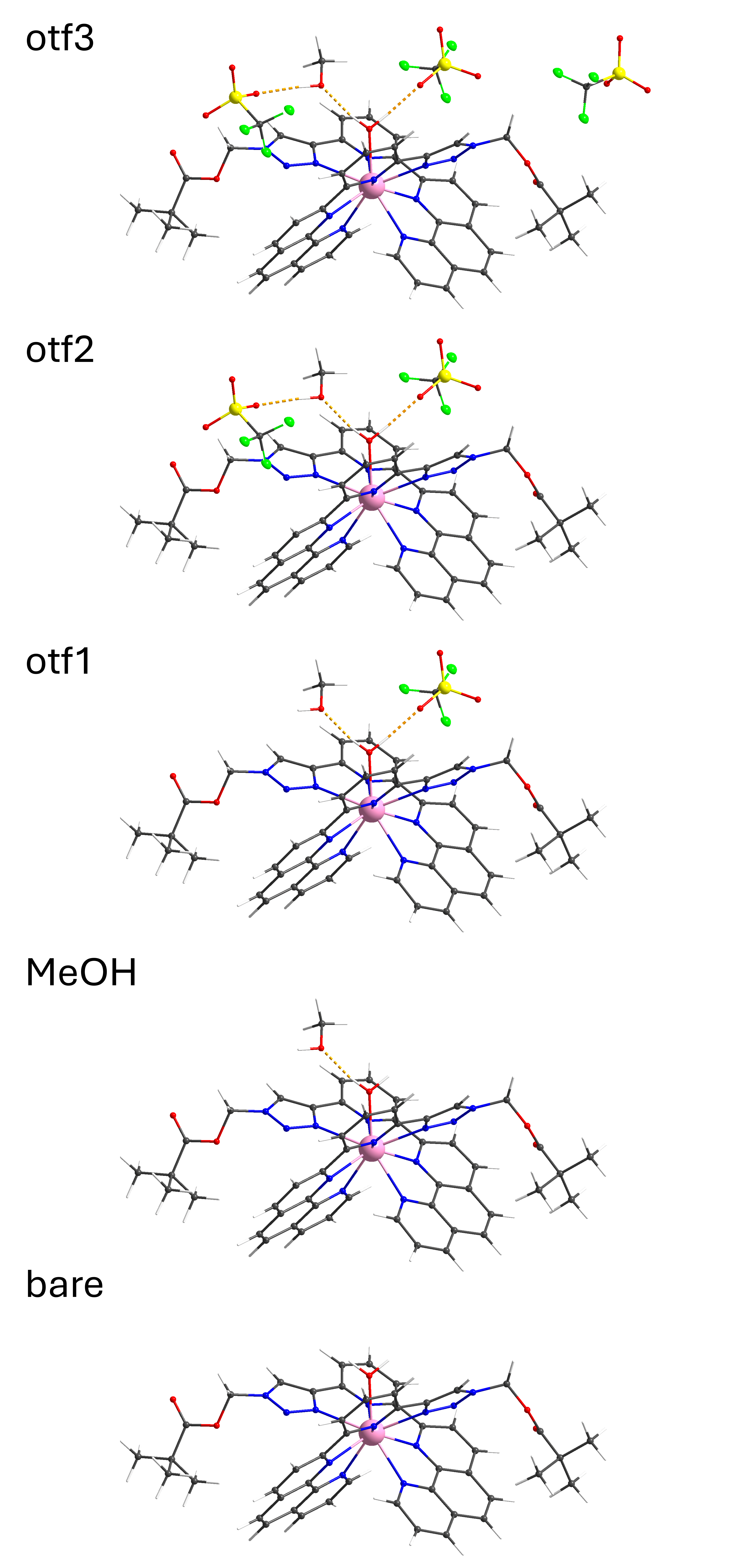}
    \caption{A representation of the five models used in this study. The otf3 model is charge neutral and contains three triflate anions. For the model otf2, the triflate furthest from the Er-centre is removed, leaving a mono-cationic complex. Next, the triflate which is H-bonded to the methanol molecule is removed, leaving a di-cationic complex otf1. In the fourth calculation, the last triflate anion is removed and the complex has a charge +3 and is named MeOH. Finally, the MeOH molecule is removed as well leaving only the PPTMP ligand and keeping the charge of +3. This model is called bare.}
    \label{fig:models}
\end{figure}
\newpage

\begin{table}[ht]
\begin{tabular}{@{}clll@{}}
\hline
energy in $cm^{-1}$  & $g_x$    & $g_y$    & $g_z$    \\ \hline
0       & 0.464113 & 1.33494  & 12.07135 \\
24.604  & 1.78109  & 3.394893 & 11.94594 \\
86.201  & 6.276628 & 4.488298 & 0.852313 \\
117.686 & 0.95098  & 1.205286 & 13.8677  \\
190.68  & 1.115064 & 1.635626 & 9.377779 \\
232.198 & 1.893061 & 4.617316 & 8.263319 \\
287.281 & 3.13556  & 3.820199 & 8.460835 \\
329.869 & 1.521774 & 3.034892 & 13.18099 \\ \hline
\end{tabular}
\caption{Energies of the lowest $^4I_{15/2}$-ground state multiplet of compound \one using the bare molecule as a model cluster.}  
\end{table}

\begin{table}[ht]
\begin{tabular}{@{}clll@{}}
\hline
energy in $cm^{-1}$  & $g_x$    & $g_y$    & $g_z$    \\ \hline
0       & 0.742199 & 1.545679 & 12.48478 \\
21.835  & 0.911138 & 4.124384 & 11.71255 \\
91.338  & 6.216954 & 3.589997 & 0.462906 \\
113.683 & 0.802401 & 1.155125 & 13.81597 \\
190.721 & 0.690271 & 1.958725 & 10.79499 \\
235.934 & 0.027568 & 2.476183 & 7.318452 \\
291.169 & 4.84864  & 5.098171 & 6.745822 \\
323.111 & 2.722719 & 3.698815 & 11.52246 \\ \hline
\end{tabular}
\caption{Energies of the lowest $^4I_{15/2}$-ground state multiplet of compound \one using the bare molecule and the MeOH molecule H-bonded to it as a model cluster.}
\end{table}

\begin{table}[ht]
\begin{tabular}{@{}clll@{}}
\hline
energy in $cm^{-1}$  & $g_x$    & $g_y$    & $g_z$    \\ \hline
0       & 1.631398 & 2.705136 & 12.22968 \\
23.107  & 0.956058 & 2.769475 & 9.756338 \\
103.512 & 4.982802 & 3.466588 & 1.53308  \\
118.033 & 0.692166 & 0.780354 & 12.04524 \\
195.933 & 0.33995  & 1.526324 & 12.04844 \\
244.904 & 5.789047 & 4.055977 & 1.125247 \\
302.132 & 0.905727 & 3.873352 & 11.07073 \\
328.556 & 4.052207 & 5.512911 & 8.931604 \\ \hline
\end{tabular}
\caption{Energies of the lowest $^4I_{15/2}$-ground state multiplet of compound \one using the bare molecule and the MeOH molecule H-bonded to it as well as the closest triflate ion as a model cluster.}
\end{table}

\begin{table}[ht]
\begin{tabular}{@{}clll@{}}
\hline
energy in $cm^{-1}$  & $g_x$    & $g_y$    & $g_z$    \\ \hline
0       & 1.113272 & 1.979091 & 12.5744  \\
25.433  & 0.905556 & 1.705587 & 9.41756  \\
108.156 & 2.146888 & 2.454909 & 5.191053 \\
121.53  & 0.674756 & 0.853468 & 11.74836 \\
198.558 & 0.209542 & 1.164288 & 12.33081 \\
249.111 & 5.480398 & 4.099915 & 1.058192 \\
304.159 & 0.565985 & 2.857098 & 12.21493 \\
332.622 & 3.407895 & 5.381822 & 9.798135 \\ \hline
\end{tabular}
\caption{Energies of the lowest $^4I_{15/2}$-ground state multiplet of compound \one using the bare molecule and the MeOH molecule H-bonded to it as well as the two closest triflate ionas a model cluster.}
\end{table}

\begin{table}[ht]
\begin{tabular}{@{}clll@{}}
\hline
energy in $cm^{-1}$  & $g_x$    & $g_y$    & $g_z$    \\ \hline
0       & 1.099575 & 1.954822 & 12.61568 \\
26.275  & 0.552248 & 1.986302 & 9.15806  \\
109.434 & 2.115618 & 2.258052 & 5.415104 \\
122.846 & 0.751401 & 0.80461  & 11.43809 \\
199.488 & 0.246097 & 1.114088 & 12.42371 \\
250.93  & 5.330429 & 4.102663 & 0.857963 \\
304.309 & 0.536737 & 2.562988 & 12.50193 \\
334.148 & 3.197615 & 5.552147 & 9.932153 \\ \hline
\end{tabular}
\caption{Energies of the lowest $^4I_{15/2}$-ground state multiplet of compound \one using the bare molecule and the MeOH molecule H-bonded to it as well as the three closest triflate ion as a model cluster.}
\end{table}

\begin{table}[ht]
\begin{tabular}{@{}clll@{}}
\hline
energy in $cm^{-1}$  & $g_x$    & $g_y$    & $g_z$    \\ \hline
0       & 0.437315 & 1.224908 & 12.14806 \\
25.33   & 1.705553 & 3.337955 & 11.90137 \\
86.709  & 6.264674 & 4.482181 & 0.724821 \\
118.019 & 0.951084 & 1.243748 & 13.78266 \\
190.987 & 1.110709 & 1.638579 & 9.502643 \\
233.377 & 1.897018 & 4.637909 & 8.1871   \\
288.522 & 3.285838 & 3.769121 & 8.383363 \\
330.262 & 1.571389 & 3.141852 & 13.07045 \\ \hline
\end{tabular}
\caption{Energies of the lowest $^4I_{15/2}$-ground state multiplet of compound \one using the bare molecule as a model cluster. A CASSCF optimization was carried out between the ROHF and CASOCI step.}
\end{table}

\begin{table}[ht]
\begin{tabular}{@{}clll@{}}
\hline
energy in $cm^{-1}$  & $g_x$    & $g_y$    & $g_z$    \\ \hline
0       & 0.856189 & 1.526494 & 12.52223 \\
22.428  & 0.798786 & 4.011603 & 11.66128 \\
92.197  & 6.247546 & 3.438501 & 0.59067  \\
113.384 & 0.822889 & 1.219776 & 13.62486 \\
190.708 & 0.673073 & 1.912892 & 10.97049 \\
236.903 & 0.052488 & 2.441664 & 7.158722 \\
292.189 & 4.531956 & 5.517375 & 6.55688  \\
322.949 & 2.97856  & 3.991183 & 11.0749  \\ \hline
\end{tabular}
\caption{Energies of the lowest $^4I_{15/2}$-ground state multiplet of compound \one using the bare molecule and the MeOH molecule H-bonded to it as a model cluster. A CASSCF optimization was carried out between the ROHF and CASOCI step.}
\end{table}

\begin{table}[ht]
\begin{tabular}{@{}clll@{}}
\hline
energy in $cm^{-1}$  & $g_x$    & $g_y$    & $g_z$    \\ \hline
0       & 1.493303 & 2.523566 & 12.36575 \\
24.725  & 0.837243 & 2.576931 & 9.778527 \\
105.373 & 1.309957 & 2.966753 & 5.391502 \\
118.907 & 0.59904  & 0.791178 & 11.15819 \\
196.364 & 0.340987 & 1.443163 & 12.1798  \\
247.391 & 5.58718  & 4.28406  & 1.32184  \\
303.162 & 0.635504 & 3.215362 & 11.95843 \\
331.352 & 3.686098 & 5.151231 & 9.798212 \\ \hline
\end{tabular}
\caption{Energies of the lowest $^4I_{15/2}$-ground state multiplet of compound \one using the bare molecule and the MeOH molecule H-bonded to it as well as the closest triflate ion as a model cluster. A CASSCF optimization was carried out between the ROHF and CASOCI step.}
\end{table}

\begin{table}[ht]
\begin{tabular}{@{}clll@{}}
\hline
energy in $cm^{-1}$  & $g_x$    & $g_y$    & $g_z$    \\ \hline
0       & 0.96626  & 1.729378 & 12.72756 \\
27.519  & 0.695202 & 1.541026 & 9.43105  \\
110.685 & 1.450073 & 2.415749 & 5.644436 \\
123.261 & 0.622711 & 0.832885 & 10.86447 \\
199.404 & 0.190473 & 1.061793 & 12.47081 \\
252.422 & 5.250636 & 4.369285 & 1.288032 \\
305.547 & 0.376178 & 2.301162 & 12.94375 \\
336.981 & 2.868193 & 5.302695 & 10.58111 \\ \hline
\end{tabular}
\caption{Energies of the lowest $^4I_{15/2}$-ground state multiplet of compound \one using the bare molecule and the MeOH molecule H-bonded to it as well as the two closest triflate ionas a model cluster. A CASSCF optimization was carried out between the ROHF and CASOCI step.}
\end{table}

\begin{table}[ht]
\begin{tabular}{@{}clll@{}}
\hline
energy in $cm^{-1}$  & $g_x$    & $g_y$    & $g_z$    \\ \hline
0       & 0.991976 & 1.510749 & 11.85706 \\
25.717  & 1.772279 & 2.953194 & 11.74088 \\
86.571  & 6.170677 & 4.82162  & 0.979597 \\
120.133 & 1.03349  & 1.249405 & 13.71368 \\
191.411 & 0.89937  & 1.712352 & 9.45204  \\
233.691 & 2.029505 & 4.89248  & 8.152496 \\
290.152 & 3.120746 & 3.417466 & 8.539775 \\
332.964 & 1.547915 & 3.254118 & 13.0068  \\ \hline
\end{tabular}
\caption{Energies of the lowest $^4I_{15/2}$-ground state multiplet of compound \one using the bare molecule as a model cluster. No optimization of H-positions was carried out.}
\end{table}

\begin{table}[ht]
\begin{tabular}{@{}clll@{}}
\hline
energy in $cm^{-1}$  & $g_x$    & $g_y$    & $g_z$    \\ \hline
0       & 0.095125 & 0.892639 & 12.42285 \\
23.467  & 1.389848 & 3.863905 & 11.88953 \\
89.222  & 6.357639 & 3.923672 & 0.02646  \\
116.299 & 0.87636  & 1.25705  & 13.73171 \\
190.634 & 0.84461  & 1.794934 & 10.57859 \\
236.396 & 0.800113 & 3.358106 & 7.443024 \\
291.657 & 4.529407 & 4.878332 & 6.943376 \\
326.758 & 2.283571 & 3.644    & 11.99173 \\ \hline
\end{tabular}
\caption{Energies of the lowest $^4I_{15/2}$-ground state multiplet of compound \one using the bare molecule and the MeOH molecule H-bonded to it as a model cluster. No optimization of H-positions was carried out.}
\end{table}

\begin{table}[ht]
\begin{tabular}{@{}clll@{}}
\hline
energy in $cm^{-1}$  & $g_x$    & $g_y$    & $g_z$    \\ \hline
0       & 1.614803 & 1.865976 & 12.46999 \\
22.86   & 0.157326 & 3.251541 & 10.70548 \\
98.265  & 5.460793 & 3.838279 & 1.238611 \\
116.995 & 0.678479 & 1.117233 & 13.12727 \\
193.617 & 0.514259 & 1.570941 & 11.80541 \\
243.106 & 6.201738 & 3.39482  & 0.539938 \\
300.819 & 2.12197  & 5.222067 & 8.691719 \\
327.023 & 7.639326 & 6.65206  & 4.028478 \\ \hline
\end{tabular}
\caption{Energies of the lowest $^4I_{15/2}$-ground state multiplet of compound \one using the bare molecule and the MeOH molecule H-bonded to it as well as the closest triflate ion as a model cluster. No optimization of H-positions was carried out.}
\end{table}

\begin{table}[ht]
\begin{tabular}{@{}clll@{}}
\hline
energy in $cm^{-1}$  & $g_x$    & $g_y$    & $g_z$    \\ \hline
0       & 1.339302 & 1.739905 & 12.59215 \\
24.068  & 0.605552 & 2.224844 & 10.24775 \\
102.075 & 4.786978 & 3.49886  & 2.08559  \\
118.859 & 0.880616 & 1.024245 & 13.20331 \\
195.313 & 0.39787  & 1.277172 & 12.08874 \\
246.24  & 0.02878  & 2.964646 & 5.982694 \\
302.587 & 1.728306 & 4.230371 & 9.76917  \\
328.713 & 7.170646 & 6.549741 & 4.62294  \\\hline
\end{tabular}
\caption{Energies of the lowest $^4I_{15/2}$-ground state multiplet of compound \one using the bare molecule and the MeOH molecule H-bonded to it as well as the two closest triflate ionas a model cluster. No optimization of H-positions was carried out.}
\end{table}

\begin{table}[ht]
\begin{tabular}{@{}clll@{}}
\hline
energy in $cm^{-1}$  & $g_x$    & $g_y$    & $g_z$    \\ \hline
0       & 1.328328 & 1.783609 & 12.5768  \\
24.455  & 0.445854 & 2.339629 & 9.926323 \\
103.128 & 4.886424 & 3.790953 & 1.789334 \\
119.945 & 0.750364 & 1.160636 & 12.96313 \\
196.235 & 0.42971  & 1.200069 & 12.19569 \\
247.535 & 5.814669 & 3.187008 & 0.010047 \\
303.029 & 1.555081 & 3.733499 & 10.33218 \\
329.973 & 7.616898 & 6.248161 & 4.591696 \\ \hline
\end{tabular}
\caption{Energies of the lowest $^4I_{15/2}$-ground state multiplet of compound \one using the bare molecule and the MeOH molecule H-bonded to it as well as the three closest triflate ion as a model cluster. No optimization of H-positions was carried out.}
\end{table}

\begin{table}[ht]
\begin{tabular}{llllll}
\hline
        & bare             & MeOH             & otf1             & otf2             & otf3             \\ \hline
B(2,-2) & 1.391387931E-01  & 1.471725309E-01  & 1.095681622E-01  & 1.361509565E-01  & 1.007382274E-01  \\
B(2,-1) & -7.317330528E-01 & -7.014531087E-01 & -6.507781407E-01 & -5.050233498E-01 & -4.753978122E-01 \\
B(2, 0) & -1.563935728E-03 & -1.811445843E-02 & -5.939380643E-02 & -7.449722337E-02 & -8.287648107E-02 \\
B(2, 1) & -5.756200913E-01 & 4.030153905E-02  & 7.858176681E-01  & 9.122341180E-01  & 9.292325241E-01  \\
B(2, 2) & -2.577715889E-01 & -2.438922860E-02 & 1.880796660E-01  & 1.642826014E-01  & 1.402323629E-01  \\
B(4,-4) & 2.464661857E-03  & 2.479978921E-03  & 2.900623710E-03  & 2.855724175E-03  & 2.980857810E-03  \\
B(4,-3) & -5.187147318E-03 & -5.329972523E-03 & -3.989113282E-03 & -5.343696334E-03 & -4.765103009E-03 \\
B(4,-2) & 2.864366640E-03  & 3.007335893E-03  & 3.785291843E-03  & 3.883190069E-03  & 4.053989911E-03  \\
B(4,-1) & -3.686727436E-03 & -3.645146768E-03 & -3.818749612E-03 & -3.876087025E-03 & -3.946742993E-03 \\
B(4, 0) & -1.561310627E-03 & -1.582881570E-03 & -1.574482520E-03 & -1.597217504E-03 & -1.554972825E-03 \\
B(4, 1) & -1.077200822E-02 & -1.103490942E-02 & -1.062382502E-02 & -1.043336077E-02 & -1.038856778E-02 \\
B(4, 2) & 4.471229684E-03  & 4.633219638E-03  & 4.932917731E-03  & 4.820056585E-03  & 4.874196131E-03  \\
B(4, 3) & 1.143558902E-02  & 1.360135435E-02  & 1.611675817E-02  & 1.682608268E-02  & 1.711218793E-02  \\
B(4, 4) & -5.494035013E-03 & -5.014160908E-03 & -4.615736898E-03 & -3.970450293E-03 & -4.100563475E-03 \\
B(6,-6) & -3.153291213E-05 & -4.485535407E-05 & -6.767281477E-05 & -7.821815746E-05 & -8.048390438E-05 \\
B(6,-5) & 1.729129924E-03  & 1.714590124E-03  & 1.811244559E-03  & 1.722334656E-03  & 1.782773184E-03  \\
B(6,-4) & -3.590098293E-04 & -3.645941475E-04 & -3.620061451E-04 & -3.686694417E-04 & -3.668822265E-04 \\
B(6,-3) & 4.189096310E-04  & 3.953942358E-04  & 3.604163606E-04  & 3.390158577E-04  & 3.373146298E-04  \\
B(6,-2) & -2.634091073E-04 & -2.567700525E-04 & -2.402816280E-04 & -2.439458946E-04 & -2.383037027E-04 \\
B(6,-1) & 3.167730154E-04  & 3.315924688E-04  & 3.511025902E-04  & 3.783557445E-04  & 3.739886895E-04  \\
B(6, 0) & 3.605547038E-05  & 3.720263969E-05  & 3.888796522E-05  & 3.890937515E-05  & 3.943248629E-05  \\
B(6, 1) & -5.222560654E-05 & -5.201318971E-05 & -5.576681047E-05 & -5.803390582E-05 & -5.766943125E-05 \\
B(6, 2) & -2.777859272E-04 & -2.816999918E-04 & -2.811057846E-04 & -2.871258141E-04 & -2.850192137E-04 \\
B(6, 3) & 6.069686102E-04  & 6.030638898E-04  & 6.175649856E-04  & 6.152729208E-04  & 6.167467003E-04  \\
B(6, 4) & 5.117310144E-05  & 4.118662834E-05  & 1.739330393E-05  & 8.249773350E-06  & 4.765798082E-06  \\
B(6, 5) & -1.156586170E-04 & -1.669756630E-04 & -3.707975602E-04 & -4.148990733E-04 & -4.476168398E-04 \\
B(6, 6) & 1.664629109E-04  & 1.670298560E-04  & 1.450479113E-04  & 1.490783976E-04  & 1.407551883E-04  \\
B(8,-8) & 1.545258468E-08  & 1.636824995E-08  & 1.595600251E-08  & 1.637877235E-08  & 1.605534613E-08  \\
B(8,-7) & -1.782808010E-08 & -1.137962137E-08 & -8.439832137E-10 & -5.393098409E-09 & -3.768599095E-09 \\
B(8,-6) & -7.593833606E-09 & -7.806865911E-09 & -7.672632190E-09 & -7.495327215E-09 & -7.586122715E-09 \\
B(8,-5) & -4.652515620E-08 & -4.718219000E-08 & -5.259876801E-08 & -5.060322851E-08 & -5.336163241E-08 \\
B(8,-4) & -1.366662765E-08 & -1.402190770E-08 & -1.637919979E-08 & -1.497117533E-08 & -1.619108702E-08 \\
B(8,-3) & -3.559754698E-08 & -3.736359334E-08 & -4.956826973E-08 & -4.686765304E-08 & -4.924346283E-08 \\
B(8,-2) & -7.883689701E-10 & -7.867133612E-09 & -1.696034288E-08 & -1.711544582E-08 & -1.733584573E-08 \\
B(8,-1) & 8.955458594E-11  & 2.054767215E-09  & 3.938586809E-09  & 3.507151480E-09  & 3.269353052E-09  \\
B(8, 0) & 5.553953168E-10  & 5.646684146E-10  & 6.421775836E-10  & 6.509963022E-10  & 6.604753092E-10  \\
B(8, 1) & -4.251213801E-09 & -1.146458902E-09 & 2.631325953E-09  & 3.312236231E-09  & 3.576480681E-09  \\
B(8, 2) & -4.224774361E-09 & -5.765977245E-09 & -5.028138078E-09 & -5.052743560E-09 & -4.473821287E-09 \\
B(8, 3) & -4.742325566E-08 & -4.533858357E-08 & -4.160339940E-08 & -3.226923969E-08 & -3.468916841E-08 \\
B(8, 4) & -2.393968587E-08 & -2.173144638E-08 & -1.843708840E-08 & -1.967696749E-08 & -1.965637171E-08 \\
B(8, 5) & 1.263044680E-08  & 1.101630984E-08  & 1.066891977E-08  & 1.691337528E-08  & 1.752795805E-08  \\
B(8, 6) & 2.371421307E-08  & 2.467738966E-08  & 2.523056487E-08  & 2.509560399E-08  & 2.489102383E-08  \\
B(8, 7) & -7.365723795E-08 & -7.366664938E-08 & -7.688628896E-08 & -8.032769695E-08 & -8.073694572E-08 \\
B(8, 8) & -2.954651783E-08 & -2.893546432E-08 & -2.918349226E-08 & -2.932420579E-08 & -2.949452050E-08 \\ \hline
\end{tabular}
\caption{Extended Stevens Operator Equivalents B(k,q) for different models of compound \one using ROHF to optimize the orbitals.}
\end{table}
\begin{table}[ht]
\begin{tabular}{lllll}
\hline
        & bare             & MeOH             & otf1             & otf2             \\ \hline
B(2,-2) & 1.439020816E-01  & 1.427131875E-01  & 1.043011046E-01  & 1.317047797E-01  \\
B(2,-1) & -7.381846977E-01 & -7.049115386E-01 & -6.612618482E-01 & -5.049609665E-01 \\
B(2, 0) & -8.775700956E-03 & -2.646070967E-02 & -6.947920827E-02 & -8.754914210E-02 \\
B(2, 1) & -5.566424875E-01 & 9.795527777E-02  & 8.940958819E-01  & 1.031519519E+00  \\
B(2, 2) & -2.692282033E-01 & -2.506149270E-02 & 1.950575866E-01  & 1.670297763E-01  \\
B(4,-4) & 2.517640742E-03  & 2.514193168E-03  & 2.933976884E-03  & 2.899342142E-03  \\
B(4,-3) & -5.196454121E-03 & -5.391131586E-03 & -4.118753263E-03 & -5.368125346E-03 \\
B(4,-2) & 2.859236867E-03  & 3.005201431E-03  & 3.803942610E-03  & 3.930378722E-03  \\
B(4,-1) & -3.710766818E-03 & -3.669453784E-03 & -3.843312160E-03 & -3.923544641E-03 \\
B(4, 0) & -1.577920431E-03 & -1.592205321E-03 & -1.580402023E-03 & -1.609946244E-03 \\
B(4, 1) & -1.084122905E-02 & -1.111597059E-02 & -1.069356464E-02 & -1.052660052E-02 \\
B(4, 2) & 4.513162734E-03  & 4.673482334E-03  & 4.990953542E-03  & 4.905085381E-03  \\
B(4, 3) & 1.136431016E-02  & 1.365442408E-02  & 1.614380978E-02  & 1.700547057E-02  \\
B(4, 4) & -5.509890799E-03 & -5.047735706E-03 & -4.655739690E-03 & -3.966080252E-03 \\
B(6,-6) & -3.257239042E-05 & -4.670750372E-05 & -6.892294228E-05 & -8.031040685E-05 \\
B(6,-5) & 1.735581972E-03  & 1.717872409E-03  & 1.818895855E-03  & 1.726082225E-03  \\
B(6,-4) & -3.593667460E-04 & -3.644790231E-04 & -3.616390645E-04 & -3.674476376E-04 \\
B(6,-3) & 4.127782809E-04  & 3.911553992E-04  & 3.564827025E-04  & 3.328487336E-04  \\
B(6,-2) & -2.640644418E-04 & -2.562959365E-04 & -2.405935463E-04 & -2.436481798E-04 \\
B(6,-1) & 3.213695373E-04  & 3.337268704E-04  & 3.543063801E-04  & 3.830724831E-04  \\
B(6, 0) & 3.570507397E-05  & 3.710559163E-05  & 3.889770770E-05  & 3.918405657E-05  \\
B(6, 1) & -5.559288353E-05 & -5.239066265E-05 & -5.752010045E-05 & -5.931376894E-05 \\
B(6, 2) & -2.791233628E-04 & -2.822747461E-04 & -2.815329776E-04 & -2.889394017E-04 \\
B(6, 3) & 6.065857473E-04  & 6.025281860E-04  & 6.171983316E-04  & 6.186305306E-04  \\
B(6, 4) & 5.010767074E-05  & 4.099992188E-05  & 1.753034470E-05  & 7.688832909E-06  \\
B(6, 5) & -1.228727593E-04 & -1.716242533E-04 & -3.781957841E-04 & -4.306128452E-04 \\
B(6, 6) & 1.664793563E-04  & 1.659619873E-04  & 1.439421241E-04  & 1.487510890E-04  \\
B(8,-8) & 1.442025508E-08  & 1.514733280E-08  & 1.480906183E-08  & 1.513828632E-08  \\
B(8,-7) & -1.820357583E-08 & -1.064933266E-08 & 8.507782132E-11  & -5.985800232E-09 \\
B(8,-6) & -7.518793359E-09 & -7.731237453E-09 & -7.427352349E-09 & -7.167702690E-09 \\
B(8,-5) & -3.434894803E-08 & -3.556138471E-08 & -4.185872863E-08 & -3.976357791E-08 \\
B(8,-4) & -1.394972072E-08 & -1.429061028E-08 & -1.656469180E-08 & -1.523944100E-08 \\
B(8,-3) & -2.836880898E-08 & -3.086110162E-08 & -4.308282621E-08 & -4.023654402E-08 \\
B(8,-2) & 1.154615445E-10  & -7.488968019E-09 & -1.682032457E-08 & -1.674498547E-08 \\
B(8,-1) & 3.772614718E-11  & 2.083657949E-09  & 4.038584881E-09  & 3.522272696E-09  \\
B(8, 0) & 5.561100334E-10  & 5.627014141E-10  & 6.479925062E-10  & 6.563931144E-10  \\
B(8, 1) & -4.287453957E-09 & -9.196019999E-10 & 3.016518581E-09  & 3.828049752E-09  \\
B(8, 2) & -4.716556486E-09 & -6.188576787E-09 & -5.631446394E-09 & -5.515007260E-09 \\
B(8, 3) & -4.377690841E-08 & -4.291291894E-08 & -3.924981605E-08 & -2.983732999E-08 \\
B(8, 4) & -2.345381575E-08 & -2.080908826E-08 & -1.703775292E-08 & -1.825728763E-08 \\
B(8, 5) & 1.013448192E-08  & 8.278087470E-09  & 7.553436793E-09  & 1.216376168E-08  \\
B(8, 6) & 2.254963090E-08  & 2.338263514E-08  & 2.375622888E-08  & 2.352575915E-08  \\
B(8, 7) & -6.775200411E-08 & -6.884357787E-08 & -7.120423021E-08 & -7.570582478E-08 \\
B(8, 8) & -2.693493940E-08 & -2.638254051E-08 & -2.672477419E-08 & -2.674020318E-08 \\ \hline
\end{tabular}
\caption{Extended Stevens Operator Equivalents B(k,q) for different models of compound \one using ROHF and CASSCF to optimize the orbitals.}
\end{table}
\begin{table}[ht]
\begin{tabular}{llllll}
\hline
        & bare             & MeOH             & otf1             & otf2             & otf3             \\ \hline
B(2,-2) & 9.345294641E-02  & 1.039931847E-01  & 4.035771658E-02  & 6.655263552E-02  & 2.969456547E-02  \\
B(2,-1) & -7.671929718E-01 & -7.432301943E-01 & -7.384871748E-01 & -5.879883760E-01 & -5.625276136E-01 \\
B(2, 0) & -2.456185928E-02 & -2.966972683E-02 & -7.921615533E-02 & -9.113271509E-02 & -1.002170586E-01 \\
B(2, 1) & -6.669690808E-01 & -1.812671118E-01 & 4.262971475E-01  & 5.150471542E-01  & 5.310993426E-01  \\
B(2, 2) & -2.950345073E-01 & -1.191196373E-01 & 5.958525961E-02  & 1.831547646E-02  & -4.105112518E-03 \\
B(4,-4) & 2.304157480E-03  & 2.338945416E-03  & 2.685367304E-03  & 2.651709547E-03  & 2.763253036E-03  \\
B(4,-3) & -5.790439069E-03 & -5.739307336E-03 & -4.694187363E-03 & -5.988774345E-03 & -5.435210626E-03 \\
B(4,-2) & 2.661114626E-03  & 2.784265186E-03  & 3.472261067E-03  & 3.570296797E-03  & 3.734028522E-03  \\
B(4,-1) & -3.724266877E-03 & -3.710151246E-03 & -3.880158411E-03 & -3.940398146E-03 & -4.009654310E-03 \\
B(4, 0) & -1.593695114E-03 & -1.610787127E-03 & -1.587405068E-03 & -1.611863866E-03 & -1.568430786E-03 \\
B(4, 1) & -1.112169383E-02 & -1.129300424E-02 & -1.092116468E-02 & -1.072818435E-02 & -1.067859529E-02 \\
B(4, 2) & 4.557298734E-03  & 4.685078882E-03  & 4.922695077E-03  & 4.808646956E-03  & 4.859645140E-03  \\
B(4, 3) & 1.137279327E-02  & 1.296276684E-02  & 1.524866352E-02  & 1.580093552E-02  & 1.609854792E-02  \\
B(4, 4) & -5.706872099E-03 & -5.374921099E-03 & -5.038834181E-03 & -4.429070427E-03 & -4.549619394E-03 \\
B(6,-6) & -3.260696993E-05 & -4.195543879E-05 & -6.377172874E-05 & -7.312152301E-05 & -7.554118442E-05 \\
B(6,-5) & 1.716171266E-03  & 1.708318667E-03  & 1.804940457E-03  & 1.714864268E-03  & 1.775731288E-03  \\
B(6,-4) & -3.617478134E-04 & -3.652099876E-04 & -3.649224453E-04 & -3.708644710E-04 & -3.692487161E-04 \\
B(6,-3) & 4.214230901E-04  & 4.028426269E-04  & 3.686820997E-04  & 3.491024975E-04  & 3.474279773E-04  \\
B(6,-2) & -2.634985957E-04 & -2.587691432E-04 & -2.424569420E-04 & -2.467214685E-04 & -2.411027268E-04 \\
B(6,-1) & 3.117543407E-04  & 3.241774877E-04  & 3.406820193E-04  & 3.677461504E-04  & 3.632857252E-04  \\
B(6, 0) & 3.661744881E-05  & 3.752024295E-05  & 3.924168452E-05  & 3.920399585E-05  & 3.974054323E-05  \\
B(6, 1) & -5.834638204E-05 & -5.881849637E-05 & -6.198668791E-05 & -6.442797704E-05 & -6.411549715E-05 \\
B(6, 2) & -2.847758118E-04 & -2.871711029E-04 & -2.869766818E-04 & -2.928549570E-04 & -2.908178565E-04 \\
B(6, 3) & 6.061306948E-04  & 6.036044415E-04  & 6.168044931E-04  & 6.141918155E-04  & 6.159490755E-04  \\
B(6, 4) & 5.979191704E-05  & 5.088503920E-05  & 2.954194387E-05  & 2.070036974E-05  & 1.731607452E-05  \\
B(6, 5) & -8.606559347E-05 & -1.308508938E-04 & -3.167375151E-04 & -3.565214114E-04 & -3.890931129E-04 \\
B(6, 6) & 1.666926342E-04  & 1.658720998E-04  & 1.454325199E-04  & 1.496156527E-04  & 1.413832060E-04  \\
B(8,-8) & 1.542256126E-08  & 1.608537884E-08  & 1.569887879E-08  & 1.610145123E-08  & 1.578261849E-08  \\
B(8,-7) & -1.893835257E-08 & -1.396058870E-08 & -4.724102369E-09 & -9.320637786E-09 & -7.940303278E-09 \\
B(8,-6) & -8.102542163E-09 & -8.135468896E-09 & -8.230120976E-09 & -7.932891932E-09 & -8.052740534E-09 \\
B(8,-5) & -4.949873808E-08 & -4.975637315E-08 & -5.632743082E-08 & -5.394897053E-08 & -5.666322113E-08 \\
B(8,-4) & -1.334393724E-08 & -1.382637089E-08 & -1.551013022E-08 & -1.408022816E-08 & -1.516527188E-08 \\
B(8,-3) & -3.472632890E-08 & -3.526682643E-08 & -4.664762899E-08 & -4.273129064E-08 & -4.540685974E-08 \\
B(8,-2) & 1.040005024E-09  & -4.676386148E-09 & -1.245125586E-08 & -1.208207664E-08 & -1.238109551E-08 \\
B(8,-1) & -3.377266173E-10 & 1.209177262E-09  & 2.658786253E-09  & 2.042976652E-09  & 1.830399377E-09  \\
B(8, 0) & 5.595976764E-10  & 5.734312054E-10  & 6.638558450E-10  & 6.769115683E-10  & 6.871708839E-10  \\
B(8, 1) & -4.483095606E-09 & -2.048810794E-09 & 1.194431485E-09  & 1.699787898E-09  & 1.945939754E-09  \\
B(8, 2) & -4.789026345E-09 & -6.139204686E-09 & -4.659883280E-09 & -4.651788038E-09 & -4.073980784E-09 \\
B(8, 3) & -5.350151217E-08 & -5.096976110E-08 & -5.149589693E-08 & -4.248997155E-08 & -4.516132292E-08 \\
B(8, 4) & -2.392104718E-08 & -2.226386610E-08 & -1.927414594E-08 & -2.088328805E-08 & -2.083613626E-08 \\
B(8, 5) & 1.295579890E-08  & 1.211850572E-08  & 1.203812117E-08  & 1.851868729E-08  & 1.909810988E-08  \\
B(8, 6) & 2.341215918E-08  & 2.422253310E-08  & 2.472379287E-08  & 2.456917765E-08  & 2.438046698E-08  \\
B(8, 7) & -7.413266596E-08 & -7.387258254E-08 & -7.566965208E-08 & -7.923040583E-08 & -7.927463936E-08 \\
B(8, 8) & -2.953093950E-08 & -2.912071471E-08 & -2.919271388E-08 & -2.946361750E-08 & -2.954262239E-08 \\ \hline
\end{tabular}
\caption{Extended Stevens Operator Equivalents B(k,q) for different models of compound \one using ROHF to optimize the orbitals. Here hydrogen positions were not optimized.}
\end{table}

\newpage
\begin{table}[ht]
\begin{tabular}{lllllllll}
\hline
$m_J(z)$ & KD1   & KD2   & KD3   & KD4   & KD5   & KD6   & KD7   & KD8   \\ \hline
+-7.5      & 0.287 & 0.021 & 0.003 & 0.072 & 0.354 & 0.132 & 0.027 & 0.103 \\
+-6.5      & 0.118 & 0.022 & 0.024 & 0.372 & 0.022 & 0.221 & 0.071 & 0.15  \\
+-5.5      & 0.025 & 0.03  & 0.259 & 0.347 & 0.072 & 0.025 & 0.178 & 0.064 \\
+-4.5      & 0.315 & 0.055 & 0.094 & 0.07  & 0.117 & 0.03  & 0.253 & 0.065 \\
+-3.5      & 0.047 & 0.232 & 0.356 & 0.032 & 0.015 & 0.045 & 0.059 & 0.214 \\
+-2.5      & 0.172 & 0.146 & 0.081 & 0.018 & 0.305 & 0.059 & 0.05  & 0.168 \\
+-1.5      & 0.027 & 0.239 & 0.14  & 0.049 & 0.091 & 0.058 & 0.315 & 0.082 \\
+-0.5      & 0.009 & 0.254 & 0.042 & 0.041 & 0.024 & 0.429 & 0.047 & 0.154 \\ \hline
\end{tabular}
\caption{$m_J(z)$ composition of the lowest 8 Kramers Doublets for compound \one using the bare molecule as the model compound. Orbitals are optimized with ROHF.}
\end{table}

\begin{table}[ht]
\begin{tabular}{lllllllll}
\hline
$m_J(z)$ & KD1   & KD2   & KD3   & KD4   & KD5   & KD6   & KD7   & KD8   \\ \hline
+-7.5      & 0.323 & 0.014 & 0.001 & 0.047 & 0.393 & 0.089 & 0.032 & 0.101 \\
+-6.5      & 0.091 & 0.043 & 0.028 & 0.461 & 0.015 & 0.215 & 0.014 & 0.134 \\
+-5.5      & 0.059 & 0.011 & 0.272 & 0.348 & 0.05  & 0.077 & 0.08  & 0.104 \\
+-4.5      & 0.312 & 0.086 & 0.044 & 0.038 & 0.148 & 0.061 & 0.264 & 0.046 \\
+-3.5      & 0.04  & 0.189 & 0.4   & 0.017 & 0.025 & 0.035 & 0.036 & 0.258 \\
+-2.5      & 0.144 & 0.125 & 0.098 & 0.019 & 0.303 & 0.081 & 0.087 & 0.143 \\
+-1.5      & 0.001 & 0.264 & 0.132 & 0.042 & 0.06  & 0.069 & 0.41  & 0.021 \\
+-0.5      & 0.029 & 0.268 & 0.025 & 0.028 & 0.006 & 0.373 & 0.078 & 0.193 \\ \hline
\end{tabular}
\caption{$m_J(z)$ composition of the lowest 8 Kramers Doublets for compound \one using the bare molecule and the MeOH molecule hydrogen bonded to it as the model compound. Orbitals are optimized with ROHF.}
\end{table}

\begin{table}[ht]
\begin{tabular}{lllllllll}
\hline
$m_J(z)$ & KD1   & KD2   & KD3   & KD4   & KD5   & KD6   & KD7   & KD8   \\ \hline
+-7.5      & 0.241 & 0.014 & 0.045 & 0.108 & 0.358 & 0.114 & 0.062 & 0.059 \\
+-6.5      & 0.01  & 0.038 & 0.123 & 0.58  & 0.081 & 0.002 & 0.11  & 0.056 \\
+-5.5      & 0.443 & 0.031 & 0.006 & 0.075 & 0.005 & 0.047 & 0.234 & 0.16  \\
+-4.5      & 0.067 & 0.199 & 0.221 & 0.086 & 0.116 & 0.105 & 0.194 & 0.012 \\
+-3.5      & 0.152 & 0.029 & 0.167 & 0.076 & 0.377 & 0.062 & 0.049 & 0.087 \\
+-2.5      & 0.018 & 0.32  & 0.048 & 0.003 & 0.035 & 0.362 & 0.169 & 0.045 \\
+-1.5      & 0.024 & 0.19  & 0.24  & 0.051 & 0.002 & 0.094 & 0.072 & 0.327 \\
+-0.5      & 0.046 & 0.179 & 0.15  & 0.021 & 0.026 & 0.215 & 0.11  & 0.254 \\ \hline
\end{tabular}
\caption{$m_J(z)$ composition of the lowest 8 Kramers Doublets for compound \one using the bare molecule, the MeOH molecule H bonded to it as well a the closest counter ion as the model compound. Orbitals are optimized with ROHF.}
\end{table}

\begin{table}[ht]
\begin{tabular}{lllllllll}
\hline
$m_J(z)$ & KD1   & KD2   & KD3   & KD4   & KD5   & KD6   & KD7   & KD8   \\ \hline
+-7.5      & 0.29  & 0.009 & 0.101 & 0.242 & 0.248 & 0.078 & 0.016 & 0.016 \\
+-6.5      & 0.015 & 0.029 & 0.065 & 0.188 & 0.134 & 0.032 & 0.397 & 0.14  \\
+-5.5      & 0.159 & 0.003 & 0.034 & 0.336 & 0.168 & 0.006 & 0.243 & 0.049 \\
+-4.5      & 0.428 & 0.115 & 0.036 & 0.009 & 0.22  & 0.1   & 0.089 & 0.004 \\
+-3.5      & 0.053 & 0.401 & 0.265 & 0.114 & 0.057 & 0.021 & 0.03  & 0.06  \\
+-2.5      & 0.006 & 0.298 & 0.154 & 0.071 & 0.107 & 0.344 & 0.013 & 0.006 \\
+-1.5      & 0.037 & 0.053 & 0.254 & 0.04  & 0.058 & 0.192 & 0.111 & 0.256 \\
+-0.5      & 0.013 & 0.092 & 0.091 & 0     & 0.007 & 0.227 & 0.1   & 0.469 \\ \hline
\end{tabular}
\caption{$m_J(z)$ composition of the lowest 8 Kramers Doublets for compound \one using the bare molecule, the MeOH molecule H bonded to it as well a the two closest counter ions as the model compound. Orbitals are optimized with ROHF.}
\end{table}

\begin{table}[ht]
\begin{tabular}{lllllllll}
\hline
$m_J(z)$ & KD1   & KD2   & KD3   & KD4   & KD5   & KD6   & KD7   & KD8   \\ \hline
+-7.5      & 0.307 & 0.009 & 0.118 & 0.267 & 0.217 & 0.065 & 0.008 & 0.009 \\
+-6.5      & 0.019 & 0.028 & 0.045 & 0.09  & 0.131 & 0.039 & 0.498 & 0.15  \\
+-5.5      & 0.081 & 0.002 & 0.047 & 0.361 & 0.272 & 0.004 & 0.206 & 0.026 \\
+-4.5      & 0.51  & 0.069 & 0.02  & 0.047 & 0.188 & 0.096 & 0.064 & 0.007 \\
+-3.5      & 0.031 & 0.564 & 0.207 & 0.078 & 0.011 & 0.03  & 0.022 & 0.057 \\
+-2.5      & 0.007 & 0.222 & 0.288 & 0.129 & 0.091 & 0.243 & 0.003 & 0.016 \\
+-1.5      & 0.039 & 0.034 & 0.197 & 0.026 & 0.087 & 0.341 & 0.093 & 0.184 \\
+-0.5      & 0.007 & 0.072 & 0.078 & 0.003 & 0.003 & 0.182 & 0.106 & 0.55  \\ \hline
\end{tabular}
\caption{$m_J(z)$ composition of the lowest 8 Kramers Doublets for compound \one using the bare molecule, the MeOH molecule H bonded to it as well a the three closest counter ions as the model compound. Orbitals are optimized with ROHF.}
\end{table}

\begin{table}[ht]
\begin{tabular}{lllllllll}
\hline
$m_J(z)$ & KD1   & KD2   & KD3   & KD4   & KD5   & KD6   & KD7   & KD8   \\ \hline
+-7.5      & 0.282 & 0.019 & 0.003 & 0.066 & 0.36  & 0.137 & 0.027 & 0.106 \\
+-6.5      & 0.118 & 0.022 & 0.019 & 0.406 & 0.019 & 0.214 & 0.062 & 0.138 \\
+-5.5      & 0.036 & 0.029 & 0.276 & 0.327 & 0.065 & 0.026 & 0.166 & 0.075 \\
+-4.5      & 0.326 & 0.057 & 0.072 & 0.06  & 0.123 & 0.028 & 0.266 & 0.067 \\
+-3.5      & 0.032 & 0.243 & 0.353 & 0.029 & 0.015 & 0.042 & 0.06  & 0.226 \\
+-2.5      & 0.175 & 0.124 & 0.088 & 0.019 & 0.313 & 0.061 & 0.06  & 0.159 \\
+-1.5      & 0.022 & 0.242 & 0.142 & 0.052 & 0.079 & 0.076 & 0.315 & 0.071 \\
+-0.5      & 0.009 & 0.264 & 0.046 & 0.041 & 0.026 & 0.415 & 0.043 & 0.156 \\ \hline
\end{tabular}
\caption{$m_J(z)$ composition of the lowest 8 Kramers Doublets for compound \one using the bare molecule as the model compound. Orbitals are optimized with ROHF and CASSCF.}
\end{table}

\begin{table}[ht]
\begin{tabular}{lllllllll}
\hline
$m_J(z)$ & KD1   & KD2   & KD3   & KD4   & KD5   & KD6   & KD7   & KD8   \\ \hline
+-7.5      & 0.312 & 0.013 & 0.001 & 0.036 & 0.403 & 0.095 & 0.034 & 0.106 \\
+-6.5      & 0.091 & 0.043 & 0.022 & 0.522 & 0.012 & 0.191 & 0.009 & 0.11  \\
+-5.5      & 0.09  & 0.012 & 0.289 & 0.302 & 0.036 & 0.087 & 0.058 & 0.126 \\
+-4.5      & 0.311 & 0.084 & 0.035 & 0.029 & 0.147 & 0.066 & 0.281 & 0.047 \\
+-3.5      & 0.018 & 0.202 & 0.375 & 0.014 & 0.05  & 0.033 & 0.031 & 0.276 \\
+-2.5      & 0.148 & 0.09  & 0.118 & 0.021 & 0.3   & 0.08  & 0.123 & 0.121 \\
+-1.5      & 0.001 & 0.267 & 0.13  & 0.044 & 0.043 & 0.111 & 0.393 & 0.01  \\
+-0.5      & 0.029 & 0.289 & 0.03  & 0.031 & 0.008 & 0.338 & 0.071 & 0.203 \\ \hline
\end{tabular}
\caption{$m_J(z)$ composition of the lowest 8 Kramers Doublets for compound \one using the bare molecule and the MeOH molecule hydrogen bonded to it as the model compound. Orbitals are optimized with ROHF and CASSCF.}
\end{table}

\begin{table}[ht]
\begin{tabular}{lllllllll}
\hline
$m_J(z)$ & KD1   & KD2   & KD3   & KD4   & KD5   & KD6   & KD7   & KD8   \\ \hline
+-7.5      & 0.25  & 0.013 & 0.061 & 0.111 & 0.351 & 0.105 & 0.063 & 0.046 \\
+-6.5      & 0.01  & 0.038 & 0.156 & 0.502 & 0.092 & 0.004 & 0.142 & 0.056 \\
+-5.5      & 0.42  & 0.025 & 0.01  & 0.1   & 0.009 & 0.043 & 0.267 & 0.126 \\
+-4.5      & 0.097 & 0.2   & 0.185 & 0.087 & 0.14  & 0.1   & 0.181 & 0.01  \\
+-3.5      & 0.144 & 0.042 & 0.18  & 0.107 & 0.339 & 0.059 & 0.057 & 0.072 \\
+-2.5      & 0.014 & 0.341 & 0.025 & 0.001 & 0.044 & 0.392 & 0.138 & 0.045 \\
+-1.5      & 0.024 & 0.172 & 0.241 & 0.073 & 0.003 & 0.068 & 0.063 & 0.357 \\
+-0.5      & 0.041 & 0.169 & 0.143 & 0.019 & 0.022 & 0.229 & 0.089 & 0.288 \\ \hline
\end{tabular}
\caption{$m_J(z)$ composition of the lowest 8 Kramers Doublets for compound \one using the bare molecule, the MeOH molecule H bonded to it as well a the closest counter ion as the model compound. Orbitals are optimized with ROHF and CASSCF.}
\end{table}

\begin{table}[ht]
\begin{tabular}{lllllllll}
\hline
$m_J(z)$ & KD1   & KD2   & KD3   & KD4   & KD5   & KD6   & KD7   & KD8   \\ \hline
+-7.5      & 0.305 & 0.007 & 0.129 & 0.224 & 0.241 & 0.069 & 0.014 & 0.011 \\
+-6.5      & 0.015 & 0.03  & 0.074 & 0.144 & 0.138 & 0.035 & 0.449 & 0.116 \\
+-5.5      & 0.135 & 0.002 & 0.051 & 0.338 & 0.191 & 0.006 & 0.245 & 0.032 \\
+-4.5      & 0.452 & 0.099 & 0.033 & 0.013 & 0.224 & 0.094 & 0.08  & 0.005 \\
+-3.5      & 0.043 & 0.451 & 0.224 & 0.131 & 0.042 & 0.019 & 0.033 & 0.056 \\
+-2.5      & 0.005 & 0.276 & 0.175 & 0.103 & 0.099 & 0.326 & 0.009 & 0.006 \\
+-1.5      & 0.036 & 0.046 & 0.229 & 0.045 & 0.06  & 0.241 & 0.093 & 0.25  \\
+-0.5      & 0.01  & 0.088 & 0.085 & 0.002 & 0.004 & 0.21  & 0.077 & 0.523 \\ \hline
\end{tabular}
\caption{$m_J(z)$ composition of the lowest 8 Kramers Doublets for compound \one using the bare molecule, the MeOH molecule H bonded to it as well a the two closest counter ions as the model compound. Orbitals are optimized with ROHF and CASSCF.}
\end{table}

\begin{table}[ht]
\begin{tabular}{lllllllll}
\hline
$m_J(z)$ & KD1   & KD2   & KD3   & KD4   & KD5   & KD6   & KD7   & KD8   \\ \hline
+-7.5      & 0.26  & 0.027 & 0.004 & 0.048 & 0.373 & 0.145 & 0.03  & 0.114 \\
+-6.5      & 0.115 & 0.016 & 0.004 & 0.466 & 0.004 & 0.207 & 0.057 & 0.132 \\
+-5.5      & 0.047 & 0.028 & 0.305 & 0.292 & 0.063 & 0.019 & 0.156 & 0.09  \\
+-4.5      & 0.331 & 0.064 & 0.069 & 0.045 & 0.115 & 0.023 & 0.296 & 0.058 \\
+-3.5      & 0.017 & 0.276 & 0.319 & 0.035 & 0.017 & 0.045 & 0.05  & 0.241 \\
+-2.5      & 0.202 & 0.086 & 0.09  & 0.015 & 0.339 & 0.051 & 0.058 & 0.159 \\
+-1.5      & 0.019 & 0.245 & 0.146 & 0.053 & 0.065 & 0.098 & 0.322 & 0.052 \\
+-0.5      & 0.01  & 0.258 & 0.062 & 0.047 & 0.024 & 0.412 & 0.032 & 0.155 \\ \hline
\end{tabular}
\caption{$m_J(z)$ composition of the lowest 8 Kramers Doublets for compound \one using the bare molecule as the model compound. Orbitals are optimized with ROHF. No hydrogen positions were optimized.}
\end{table}

\begin{table}[ht]
\begin{tabular}{lllllllll}
\hline
$m_J(z)$ & KD1   & KD2   & KD3   & KD4   & KD5   & KD6   & KD7   & KD8   \\ \hline
+-7.5      & 0.303 & 0.014 & 0.002 & 0.047 & 0.398 & 0.098 & 0.031 & 0.107 \\
+-6.5      & 0.099 & 0.031 & 0.012 & 0.479 & 0.01  & 0.212 & 0.022 & 0.134 \\
+-5.5      & 0.056 & 0.015 & 0.307 & 0.305 & 0.053 & 0.059 & 0.102 & 0.102 \\
+-4.5      & 0.327 & 0.075 & 0.048 & 0.041 & 0.137 & 0.045 & 0.276 & 0.051 \\
+-3.5      & 0.029 & 0.225 & 0.363 & 0.028 & 0.024 & 0.036 & 0.044 & 0.25  \\
+-2.5      & 0.162 & 0.11  & 0.097 & 0.019 & 0.31  & 0.08  & 0.073 & 0.149 \\
+-1.5      & 0.005 & 0.257 & 0.136 & 0.047 & 0.06  & 0.081 & 0.384 & 0.028 \\
+-0.5      & 0.018 & 0.272 & 0.034 & 0.034 & 0.009 & 0.387 & 0.067 & 0.178 \\ \hline
\end{tabular}
\caption{$m_J(z)$ composition of the lowest 8 Kramers Doublets for compound \one using the bare molecule and the MeOH molecule hydrogen bonded to it as the model compound. Orbitals are optimized with ROHF. No hydrogen positions were optimized.}
\end{table}

\begin{table}[ht]
\begin{tabular}{lllllllll}
\hline
$m_J(z)$ & KD1   & KD2   & KD3   & KD4   & KD5   & KD6   & KD7   & KD8   \\ \hline
+-7.5      & 0.243 & 0.011 & 0.011 & 0.018 & 0.423 & 0.123 & 0.064 & 0.108 \\
+-6.5      & 0.026 & 0.04  & 0.076 & 0.804 & 0.02  & 0.024 & 0.011 & 0.001 \\
+-5.5      & 0.449 & 0.032 & 0.082 & 0.021 & 0.003 & 0.103 & 0.065 & 0.244 \\
+-4.5      & 0.046 & 0.145 & 0.311 & 0.05  & 0.015 & 0.105 & 0.293 & 0.035 \\
+-3.5      & 0.122 & 0.108 & 0.008 & 0.004 & 0.445 & 0.067 & 0.023 & 0.222 \\
+-2.5      & 0.06  & 0.103 & 0.257 & 0.033 & 0.055 & 0.115 & 0.327 & 0.05  \\
+-1.5      & 0.015 & 0.28  & 0.117 & 0.005 & 0.015 & 0.354 & 0.091 & 0.123 \\
+-0.5      & 0.039 & 0.282 & 0.139 & 0.065 & 0.025 & 0.108 & 0.126 & 0.217 \\ \hline
\end{tabular}
\caption{$m_J(z)$ composition of the lowest 8 Kramers Doublets for compound \one using the bare molecule, the MeOH molecule H bonded to it as well a the closest counter ion as the model compound. Orbitals are optimized with ROHF. No hydrogen positions were optimized.}
\end{table}

\begin{table}[ht]
\begin{tabular}{lllllllll}
\hline
$m_J(z)$ & KD1   & KD2   & KD3   & KD4   & KD5   & KD6   & KD7   & KD8   \\ \hline
+-7.5      & 0.244 & 0.007 & 0.022 & 0.133 & 0.358 & 0.118 & 0.048 & 0.069 \\
+-6.5      & 0.002 & 0.029 & 0.058 & 0.64  & 0.092 & 0.002 & 0.104 & 0.074 \\
+-5.5      & 0.449 & 0.014 & 0.002 & 0.085 & 0.01  & 0.048 & 0.189 & 0.202 \\
+-4.5      & 0.077 & 0.222 & 0.246 & 0.031 & 0.103 & 0.105 & 0.208 & 0.009 \\
+-3.5      & 0.176 & 0.031 & 0.21  & 0.029 & 0.364 & 0.041 & 0.026 & 0.123 \\
+-2.5      & 0.009 & 0.358 & 0.032 & 0.005 & 0.031 & 0.343 & 0.19  & 0.032 \\
+-1.5      & 0.013 & 0.174 & 0.293 & 0.035 & 0.003 & 0.083 & 0.108 & 0.291 \\
+-0.5      & 0.03  & 0.166 & 0.137 & 0.044 & 0.039 & 0.259 & 0.126 & 0.199 \\ \hline
\end{tabular}
\caption*{$m_J(z)$ composition of the lowest 8 Kramers Doublets for compound \one using the bare molecule, the MeOH molecule H bonded to it as well a the two closest counter ions as the model compound. Orbitals are optimized with ROHF. No hydrogen positions were optimized.}
\end{table}

\begin{table}[ht]
\begin{tabular}{lllllllll}
\hline
$m_J(z)$ & KD1   & KD2   & KD3   & KD4   & KD5   & KD6   & KD7   & KD8   \\ \hline
+-7.5      & 0.253 & 0.008 & 0.032 & 0.182 & 0.327 & 0.108 & 0.039 & 0.051 \\
+-6.5      & 0.007 & 0.026 & 0.057 & 0.5   & 0.116 & 0.009 & 0.173 & 0.112 \\
+-5.5      & 0.365 & 0.007 & 0.014 & 0.167 & 0.041 & 0.028 & 0.22  & 0.158 \\
+-4.5      & 0.18  & 0.208 & 0.163 & 0.014 & 0.157 & 0.104 & 0.17  & 0.006 \\
+-3.5      & 0.148 & 0.105 & 0.289 & 0.043 & 0.254 & 0.031 & 0.026 & 0.103 \\
+-2.5      & 0.005 & 0.396 & 0.001 & 0.005 & 0.062 & 0.385 & 0.122 & 0.025 \\
+-1.5      & 0.016 & 0.12  & 0.33  & 0.058 & 0.007 & 0.021 & 0.135 & 0.312 \\
+-0.5      & 0.026 & 0.13  & 0.114 & 0.031 & 0.037 & 0.316 & 0.114 & 0.232 \\ \hline
\end{tabular}
\caption{$m_J(z)$ composition of the lowest 8 Kramers Doublets for compound \one using the bare molecule, the MeOH molecule H bonded to it as well a the three closest counter ions as the model compound. Orbitals are optimized with ROHF. No hydrogen positions were optimized.}
\end{table}

\begin{table}[]
\begin{tabular}{lllllll}
\hline
\multirow{2}{*}{model} & \multicolumn{3}{l}{KD1}                     & \multicolumn{3}{l}{KD2}                       \\
                       & $g_{zx}$  & $g_{zy}$  & $g_{zz}$  & $g_{zx}$ & $g_{zy}$ & $g_{zz} $           \\ \hline
unopt\_bare            & 0.5104747185 & 0.5128666679  & 0.6902052903 & -0.3213535523 & 0.7095821980  & -0.6270765493 \\
unopt\_full            & 0.5733424058 & 0.1263872376  & 0.8095089573 & -0.2904445274 & 0.7223026596  & -0.6276311372 \\
opt\_bare              & 0.5637825137 & 0.5267287951  & 0.6361651151 & -0.2490059244 & 0.7347355421  & -0.6309989958 \\
opt\_full              & 0.5945358505 & -0.0246243062 & 0.8036919597 & 0.3278757114  & -0.6472654433 & 0.6881460338  \\ \hline
\end{tabular}
\caption{Direction of $g_z$-axis for different models of the first and second KD.}
\end{table}

\begin{figure}[htb]
    \centering
    \includegraphics[width = 0.8\columnwidth]{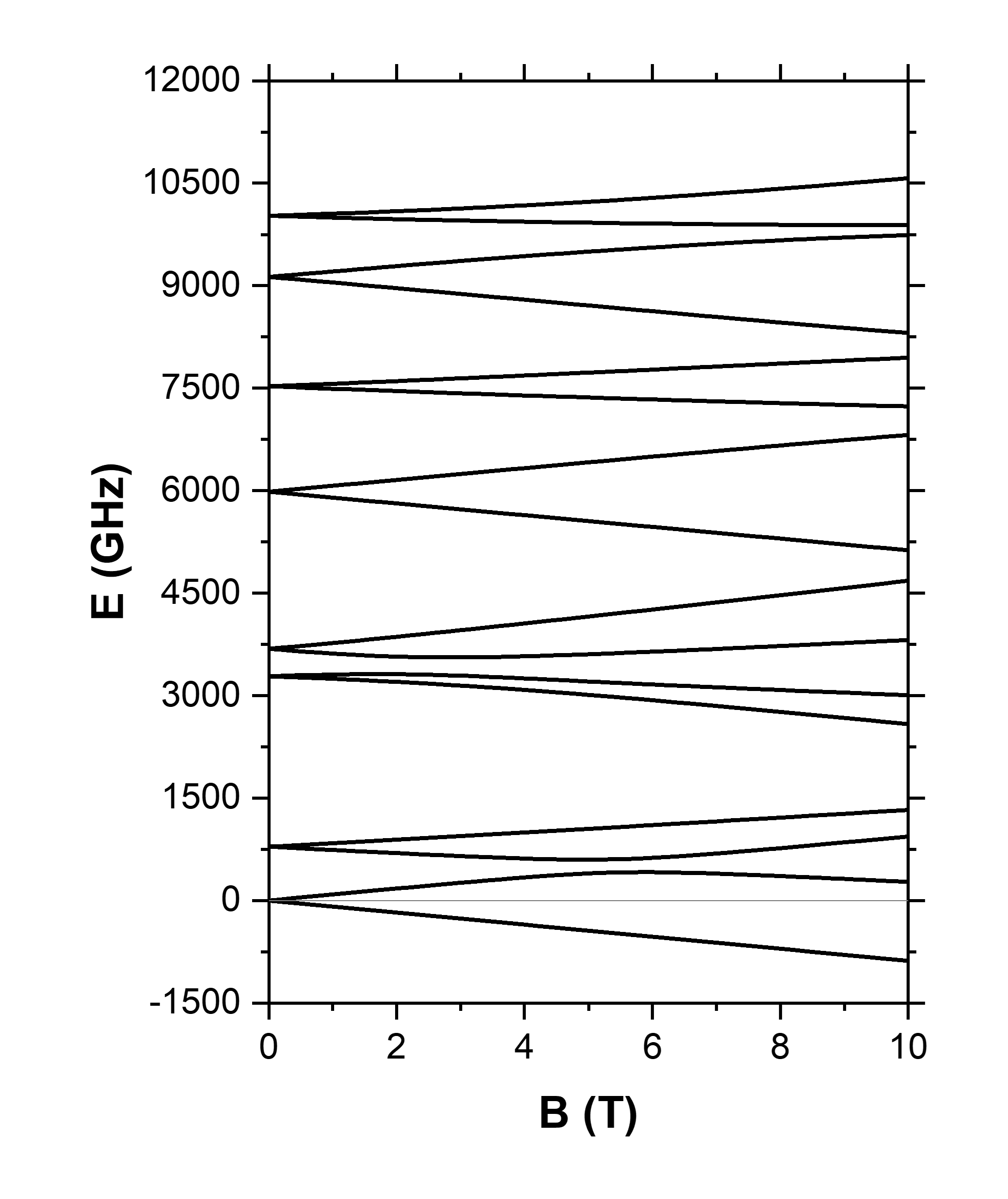}
    \caption{Zeeman diagram of all 16 states of the $^4I_{15/2}$-ground state with a magnetic field along the magnetic main axis simulating the loose powder experiment.}
    \label{fig:zeefull}
\end{figure}

\begin{figure}[htb]
    \centering
    \includegraphics[width = \columnwidth]{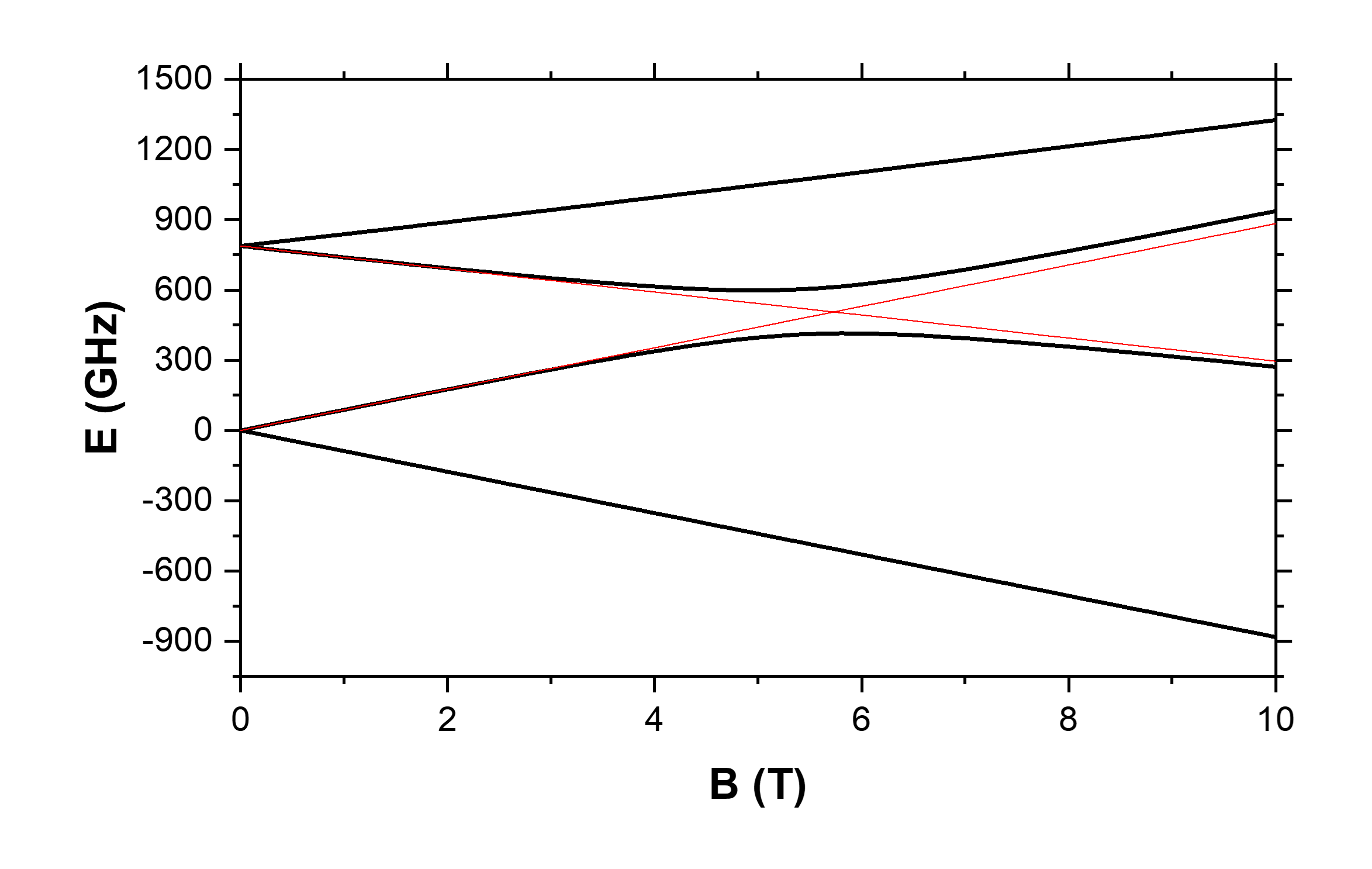}
    \caption{Zeeman diagram of the lowest four states with a magnetic field along the magnetic main axis simulating the loose powder experiment are shown in black. The red lines indicate a perfect linear behaviour, showing that a theoretical level crossing would occur at aroun 5.8 T.}
    \label{fig:zeelow}
\end{figure}

\begin{figure}[htb]
    \centering
    \includegraphics[width = \columnwidth]{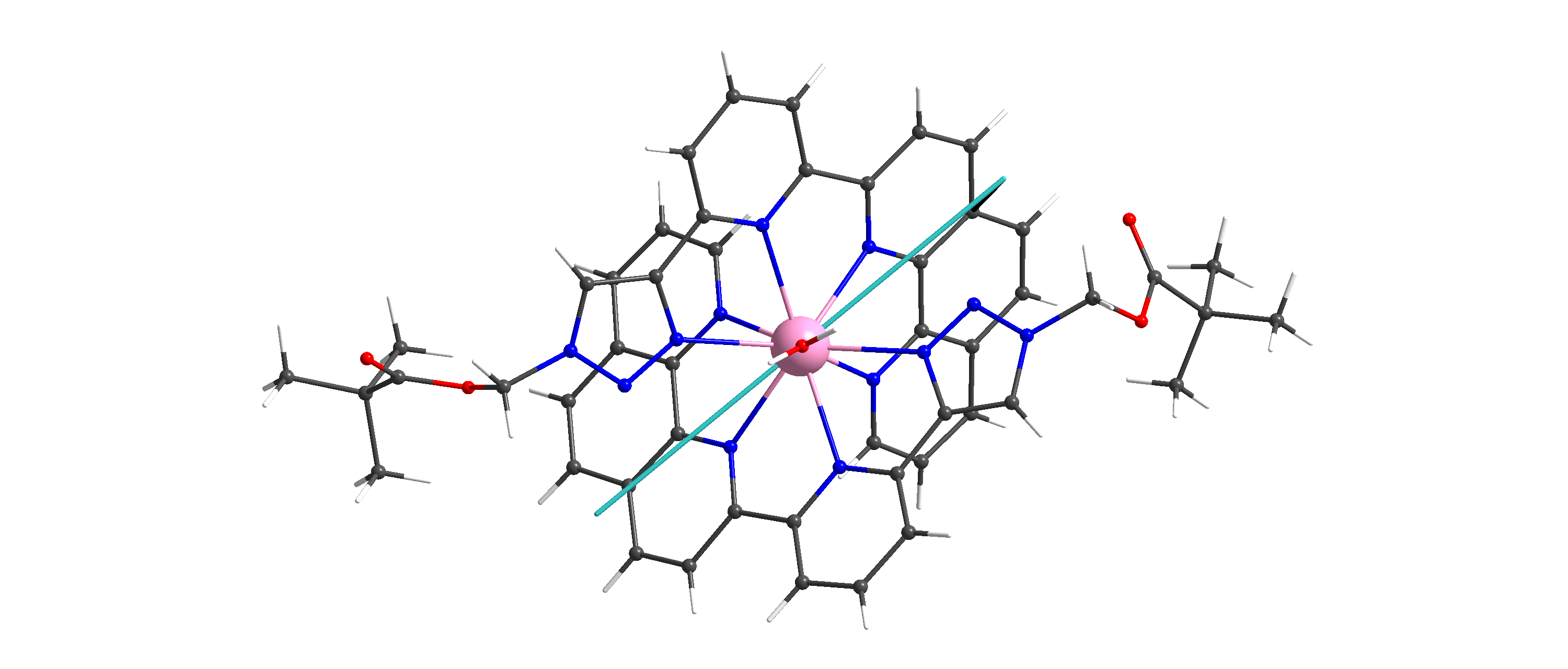}
    \caption{Calculated magnetic main axis of the ground state for the bare model with unoptimized hydrogen positions. The axis roughly follows the direction of the hydrogen positions of the water ligand.}
    \label{fig:aniso2}
\end{figure}

\begin{figure}[htb]
    \centering
    \includegraphics[width = \columnwidth]{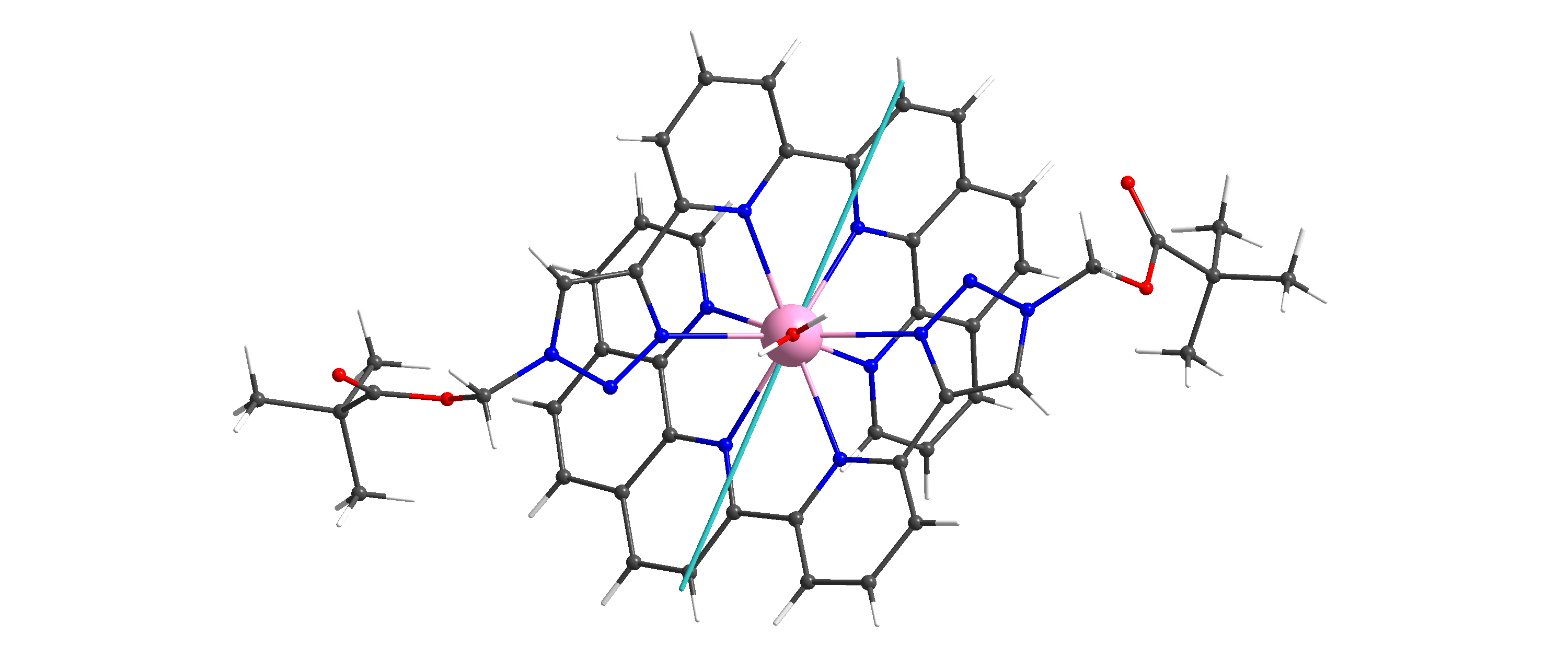}
    \caption{Calculated magnetic main axis of the ground state for the model including all counterions with unoptimized hydrogen positions. The additional MeOH and counterions are included in the calculation but omitted in this picture for simplicity.}
    \label{fig:aniso2}
\end{figure}

\begin{figure}[htb]
    \centering
    \includegraphics[width = \columnwidth]{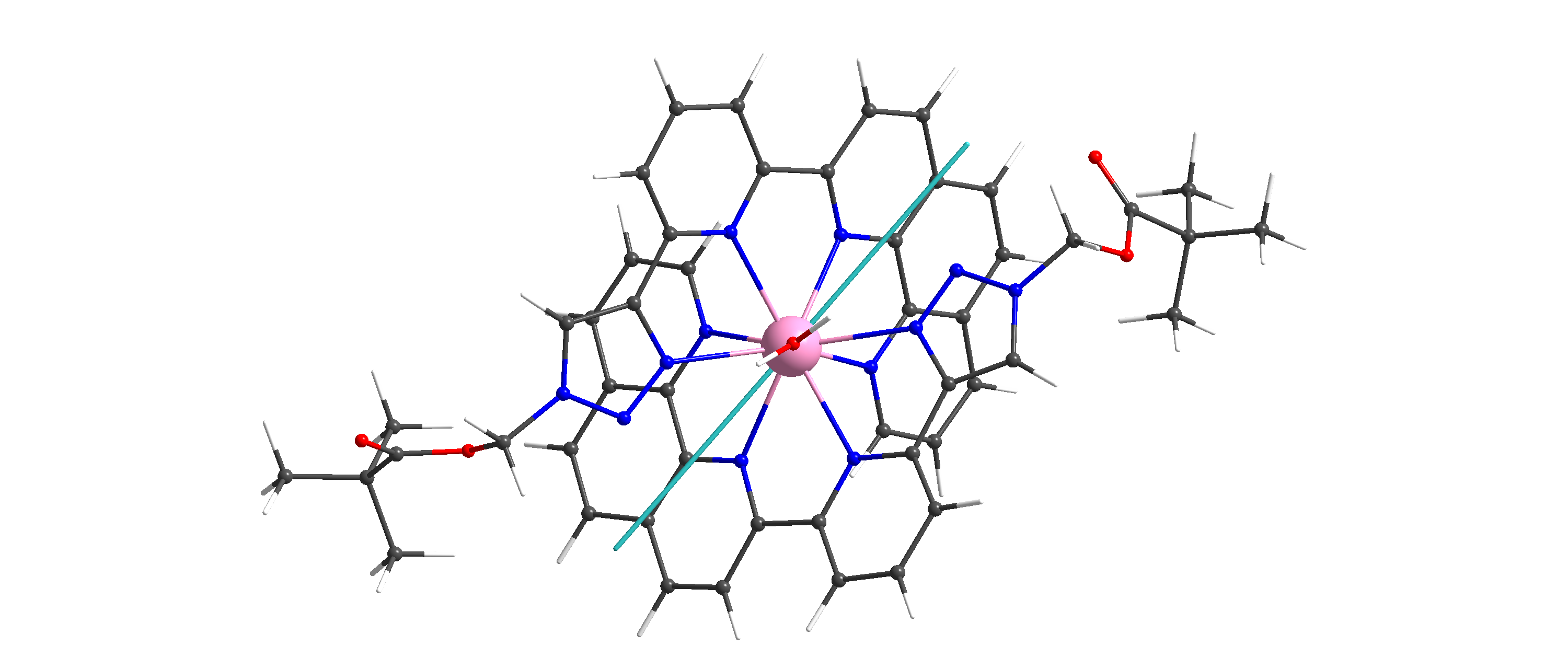}
    \caption{Calculated magnetic main axis of the ground state for the bare model with optimized hydrogen positions. The axis roughly follows the direction of the hydrogen positions of the water ligand.}
    \label{fig:aniso2}
\end{figure}

\newpage
\begin{figure}[htb]
    \centering
    \includegraphics[width = \columnwidth]{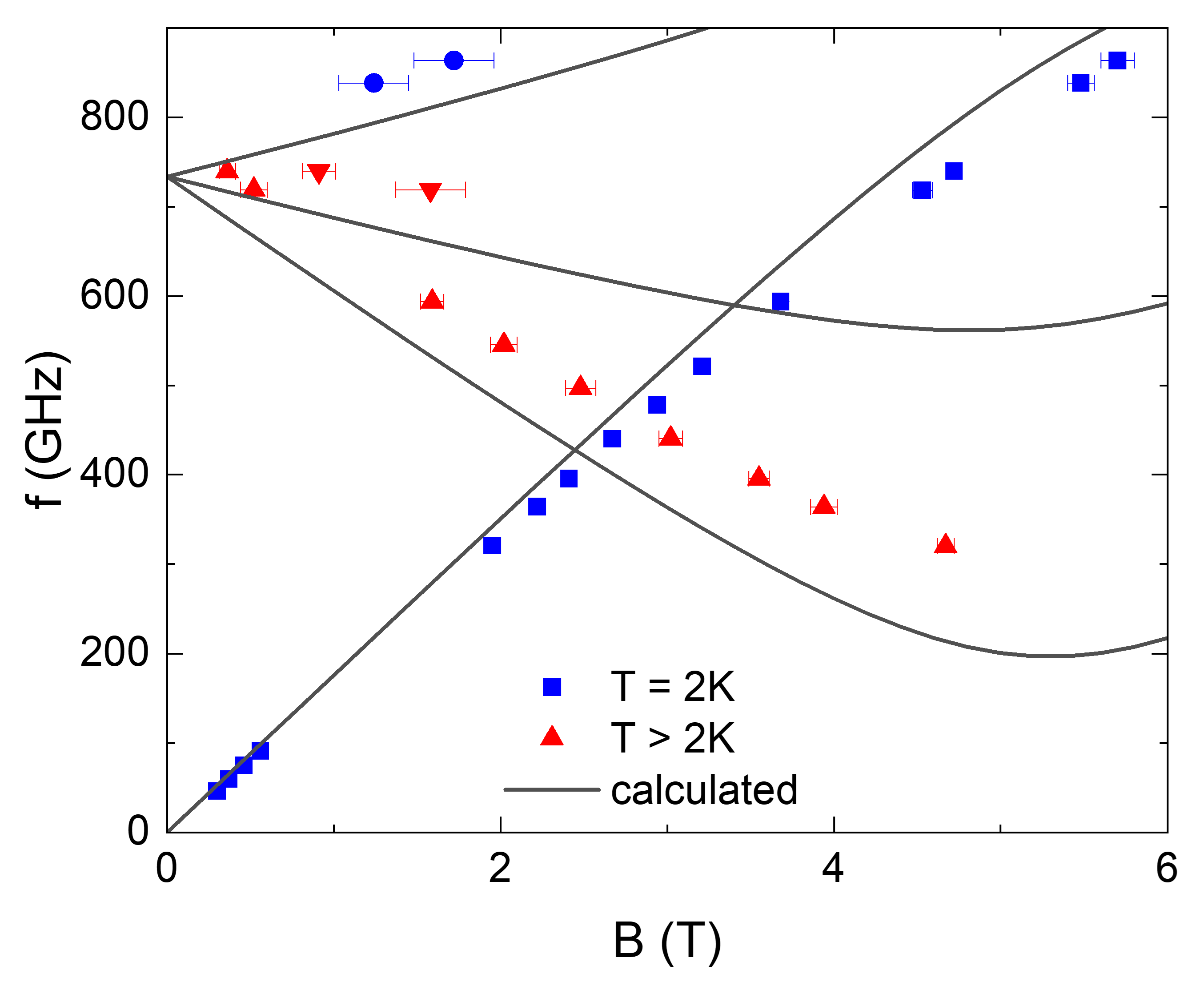}
    \caption{Measured transition energies on a loose powder (points) and calculated transition energies within the lowest two Kramers Doublets (lines) for the model containing all counterions but hydrogen positions are taken from the .cif-file.}
    \label{fig:zeesim2}
\end{figure}



\end{document}